\documentclass{article}

\usepackage{amsmath,amsthm,amssymb,epsfig,listings,bm}

\usepackage{subfigure,float}
\usepackage{graphicx}
\usepackage[dvipsnames]{xcolor}
\usepackage[font=small, labelfont=bf]{caption}
\usepackage{tcolorbox}
\usepackage{multirow}
\usepackage{array}

\usepackage{authblk}
\usepackage{blindtext}
\usepackage{multicol}

\usepackage[super,comma,sort&compress]{natbib}

\newcommand\ChangeRT[1]{\noalign{\hrule height #1}}

\newcommand{\ba}{b^{}_a}
\newcommand{\bv}{b^{}_v}
\newcommand{\rb}{r^{}_b}
\newcommand{\fb}{f^{}_b}
\newcommand{\pzero}{p^{}_0}
\newcommand{\qzero}{q^{}_0}
\newcommand{\dmax}{d^{}_{max}}
\newcommand{\xG}{x^{}_G}
\newcommand{\xD}{x^{}_D}
\newcommand{\xV}{x^{}_V}
\newcommand{\xoned}{x^{}_{1d}}
\newcommand{\xtwod}{x^{}_{2d}}

\newtcolorbox{mybox}[1]{before=\par\smallskip\centering,float=htpb!,colback=red!5!white,colframe=green!75!black,fonttitle=\bfseries,title=#1, width=1\textwidth}

\makeatletter
  \def\env@matrix{\hskip -\arraycolsep
  \let\@ifnextchar\new@ifnextchar
  \array{*\c@MaxMatrixCols l}}
\makeatother

\def\R{{\mathbb R}}

\usepackage{mwe}
\usepackage{fancyhdr}
\newlength{\tempdima}
\newcommand{\rowname}[1]
{\rotatebox{90}{\makebox[\tempdima][c]{\textbf{#1}}}}
\begin{document}

\providecommand{\keywords}[1]
{
  \small	
  \textbf{\textit{Keywords:}} #1
}

\title{Optimizing adaptive cancer therapy: dynamic programming and evolutionary game theory}
\date{\vspace{-5ex}}

\author[1]{Mark Gluzman}
\author[2]{Jacob G. Scott}
\author[,3]{Alexander Vladimirsky\footnote{Corresponding author: vladimirsky@cornell.edu}}
\affil[1]{Center for Applied Mathematics, Cornell University, Ithaca, NY}
\affil[2]{Department of Translational Hematology and Oncology Research, Cleveland Clinic, Cleveland, OH}
\affil[3]{Department of Mathematics and Center for Applied Mathematics, Cornell University, Ithaca, NY}

\pagestyle{plain}

\maketitle

\thispagestyle{fancy}

\abstract{
\textbf{BACKGROUND:}
Recent clinical trials have shown that the adaptive drug therapy
can be more efficient than a standard MTD-based policy in treatment of cancer patients.
The adaptive therapy paradigm is not based on a preset schedule; instead, the doses are administered based on the current state of tumor.
But the adaptive treatment policies examined so far have been largely \textit{ad hoc}.
In this paper we propose a method for systematically optimizing the rules of adaptive policies based on an Evolutionary Game Theory model of cancer dynamics.

\textbf{METHODS:}
Given a set of treatment objectives, we use the framework of dynamic programming to find the optimal treatment strategies.
In particular, we  optimize the total drug usage and time to recovery by solving a Hamilton-Jacobi-Bellman equation based on a mathematical model of tumor evolution.

\textbf{RESULTS:}
We compare adaptive/optimal treatment strategy with MTD-based treatment policy. We show that optimal treatment strategies can dramatically decrease the total amount of drugs prescribed as well as increase the fraction of initial tumour states from which the recovery is possible. We also examine the optimization trade-offs between the total administered drugs and recovery time.

\textbf{CONCLUSIONS:}
 The adaptive therapy combined with optimal control theory is a promising concept in the cancer treatment and should be integrated  into  clinical trial  design.
}


\section{Introduction}

Intratumoral heterogeneity is being increasingly recognized as a cause of metastasis, progression and resistance to therapy \cite{Marusyk2010}. While genetic instability, a hallmark of malignancy \cite{Hanahan2000}, can result in this heterogeneity, it is being increasingly understood that eco-evolutionary factors, like selection and clonal interference, can also drive and maintain it\cite{Scott2017, Marusyk2014}.

While sequencing technologies have enabled increasingly in-depth quantitative understanding of the genetic heterogeneity, relatively little experimental work has sought to directly quantify the eco-evolutionary interactions involved.
As more studies come to light showing the efficacy of treatments based on eco-evolutionary trial designs, this lack of quantification is coming into focus.

In line with standard, cell-autonomous growth-based theories, conventional chemotherapy is given to patients at the \emph{maximum tolerated doses} (MTD): the highest doses that most patients can safely tolerate.
Although the MTD-based chemotherapy offers advantages in survival compared to no therapy, cures remain elusive, and side effects can be severe.
In addition to the toxicity, it is known that relapse is nearly inevitable due to the emergence of therapeutic resistance: a process driven by Darwinian evolutionary dynamics in which the MTD-based chemotherapy kills off the chemotherapy-sensitive cells and chemo-refractory cells eventually dominate in the tumor.
While it is unknown whether these resistant cells are present before therapy, or acquire resistance mutations during therapy, it is the process of variation and selection under standard therapy that drives the inevitable failure in the patient.

\emph{Metronomic chemotherapy} has been proposed as a possible alternative to the MTD strategy\cite{Hanahan2000a, Lien2013}.
Metronomic chemotherapy is given in an on/off fashion at frequent time intervals according to a set periodic schedule.
The idea behind this method is to give less overall therapy, thereby increasing tolerability, and allowing time for therapy sensitive cells to regrow, allowing for resensitization of the tumor.
Frustratingly, the results of clinical trials of metronomic chemotherapy have been ``variable'' \cite{Pasquier2010} and many of them have \emph{not} demonstrated significant efficacy \cite{Kesari2007, Steinbild2007, Senerchia2017} as compared to standard therapy.
Recent studies argue that metronomic chemotherapy highly depends on timing, and the right scheduling can improve the results of its usage \cite{Chen2014}.

Based on the hypothesis that disease dynamics depend on the evolution of tumor heterogeneity as modulated by competition between subtypes, the idea of using \emph{adaptive therapy} (AT) has been proposed~\cite{Gatenby2009}.
AT is much like metronomic therapy, with an important difference.
AT administers doses of therapy according to the current \emph{state} of tumor growth and its anticipated evolutionary changes (\emph{trajectory}). These can be estimated using direct (e.g., taking biopsies) or indirect (e.g., antigen testing, mathematical modeling) methods.
Therefore, unlike the MTD-based or metronomic protocols, AT does not have preset schedule and it adjusts the doses and timing before a tumor becomes chemotherapy-resistant, prolonging time to this event.
Recently, the adaptive strategies have shown promise in pre-clinical trials of breast cancer~\cite{Enriquez-Navas2016} and a phase 2 clinical trial in metastatic castrate-resistant prostate cancer~\cite{Zhang2017a}.

These two recent successes~\cite{Enriquez-Navas2016,Zhang2017a} in adaptive therapy have been based on mathematical modeling of tumor evolution under therapy using a dynamical systems approach based on Evolutionary Game Theory (EGT) \cite{Smith1982, Hofbauer1998a}.
This formalism explicitly considers interactions between sub-populations and models their fitness in frequency dependent terms.
EGT has been used to theoretically consider many scenarios in cancer before, including therapy scheduling and timing in prostate cancer \cite{Basanta2012, You2017, Zhang2017a, Cunningham2018}; the use of tumor microenvironment targeting therapy in glioblastoma \cite{Basanta2011a}; the trade-off between healthy tissue and cancer in multiple myeloma \cite{Dingli2009, Wu2014}; and drug resistance in general \cite{Komarova2005, Orlando2012, West2016}.
These theoretical studies, combined with the recent empiric realizations, suggest significant opportunities to improve therapy by using this evolutionarily enlightened approach.
Nevertheless, therapeutic decisions in general practice are currently \emph{not} based on this knowledge and continue to use the MTD paradigm.

Assuming that an oncologist has perfect information about the current state of a tumor, and a faithful mathematical model that can predict its trajectory (these, of course, are very strong assumptions), it is not clear how he/she should adjust the schedule and doses.
Based on a stage of the disease and patient's needs, the therapy can have different final goals: maximization of patient's duration of life, ensuring the best possible quality of the rest of life, decreasing probability of new metastases appearing, decreasing time/cost of the treatment, etc..
Unfortunately, an oncologist can usually only focus on one or two of these goals, having some reasonable constraints on the secondary parameters.
Thus, an important step toward optimizing AT is to define an \emph{objective} of the therapy and ``translate'' it into mathematical language.
The next step is to \emph{quantify} how good each particular strategy is with respect to that chosen objective.
Optimizing this objective is a mathematical goal which can be addressed by the tools of optimal control theory.

Optimal control theory, a branch of mathematics typically applied to problems in engineering, can be applied to a wide class of problems arising in oncology \cite{Schattler2015}.
The first application of optimal control theory in cancer was done by Swan and Vincent~\cite{Swan1977} who found the optimal treatment strategy for multiple myeloma with the objective to minimize the total amount of drugs used applying the Pontryagin Minimum Principle (PMP)~\cite{Pontryagin1962}.
Since then, others have used the PMP to different optimal cancer treatment problems:   a chemotherapy optimization under evolving drug resistance \cite{Schattler2006, Wang2016, Carrere2017}, optimal scheduling of a vessel disruptive agent  \cite{ DOnofrio2009}, MAPK  inhibitors \cite{Su2016} input in cancer treatment, minimization  amount of drugs prescribed  in tumor¨Cimmune  model \cite{Ledzewicz2012},  finding compromise between drug toxicity and tumor repression for the myeloma bone disease \cite{Lemos2016}, and many others.

While these approaches have offered benefits in their ability to formally optimize problems written as dynamical systems, the PMP method has several limitations.
First, PMP yields only a necessary condition for an optimum, and any \emph{locally} optimal trajectory of the control system satisfies PMP.
Local optimality means that the trajectory is optimal when comparing it with its small perturbations, but there may well be a different trajectory that is even better (\emph{globally} optimal, compared with {\em all} possible trajectories).
Secondly, PMP provides a time-dependent (open-loop) control: given an initial state, the method provides an optimal treatment strategy as a function of time -- therefore a treating oncologist has to follow it regardless of the changing state of the tumor.
However, if the underlying model has been perturbed or includes some noise (like a tumor acquiring mutations, say), the control cannot adapt to these unexpected changes.

A different approach to analysis of an optimal control problem is a feedback (closed-loop) control point of view.
Using the Hamilton-Jacobi-Bellman (HJB) equations, one can obtain controls that depend on the current state of the dynamical system (current distribution of sub-populations of cancer cells) rather than only the current time\cite{Bardi1997,Liberzon2012}.
In this case, the treating oncologist's decisions can be adjusted if something unexpected has happened with the trajectory.
Moreover, the HJB equations guarantees that the resulting treatment feedback strategies are {\em globally} optimal.
Despite these advantages of the HJB over PMP method, there are few works~\cite{Nowakowski2013, Lorz2013} which use the feedback control paradigm to find an optimal treatment strategy.

Here, we apply the HJB approach to solve for optimal treatment strategies for a model of lung cancer proposed by Kaznatcheev and colleagues \cite{Kaznatcheev2017a}. In that
paper the authors introduce an evolutionary game (system of replicator-type equations) that models the dynamics of three sub-populations of tumor cells.
The article highlights the importance of a good scheduling in the polyclonal regime, when the game has cyclic dynamics.
The article has an example of two different scheduling strategies with the same set of initial parameters that lead the system to opposite outcomes: putative recovery, versus putative death of a patient.
While several qualitatively different treatment schedules are presented, optimal therapy is not discussed.
Given the growing interest in EGT in clinical applications\cite{Zhang2017a} and recent work connecting these models using direct \textit{in vitro} parameterization
\cite{Kaznatcheev2017b},
we believe the optimization of therapies based on such models will become increasingly important and the HJB-based approach will be used far more often in the future.

\section{Methods}\label{Methods}

In this article we focus on a model of cancer evolution that has been proposed in Kaznatcheev et al.\cite{Kaznatcheev2017a} and summarized in Box~1.
This model considers interactions between three different sub-populations of cancer cells playing a modified version of the public goods social dilemma: glycoltyic cells (GLY), vacscular overproducers (VOP) and cells called defectors (DEF) which use both strategies to ``cheat'' on the others.
GLY cells are anaerobic and produce lactic acid.
Both VOP and DEF cells are aerobic. In addition, VOP cells spend extra energy to produce VEGF (a protein that improves the vasculature, benefiting both VOP and DEF).
Based on the replicator model from EGT\cite{Smith1982, Hofbauer1998a} summarized in equations \eqref{re} and \eqref{re2d}, the evolution of the tumor is described by tracking the changing {\em proportions} of GLY, VOP and DEF cells in the full population.
The patient is viewed as recovered when the GLY proportion falls below some low threshold $\rb$.
(Below this {\em recovery barrier}, the validity of replicator-based model is harder to justify and we assume that the GLY cells are essentially extinct.)
Conversely, we assume that GLY cells \emph{suppress} other tumor cells and a patient dies if the total proportion of aerobic cells (VOP and DEF sub-populations combined) falls below some low threshold (a \emph{failure barrier}) $\fb$.

\begin{figure}[H]
\centering
\subfigure[][]{%
\label{fig:ex4-a}%
\includegraphics[height=3in]{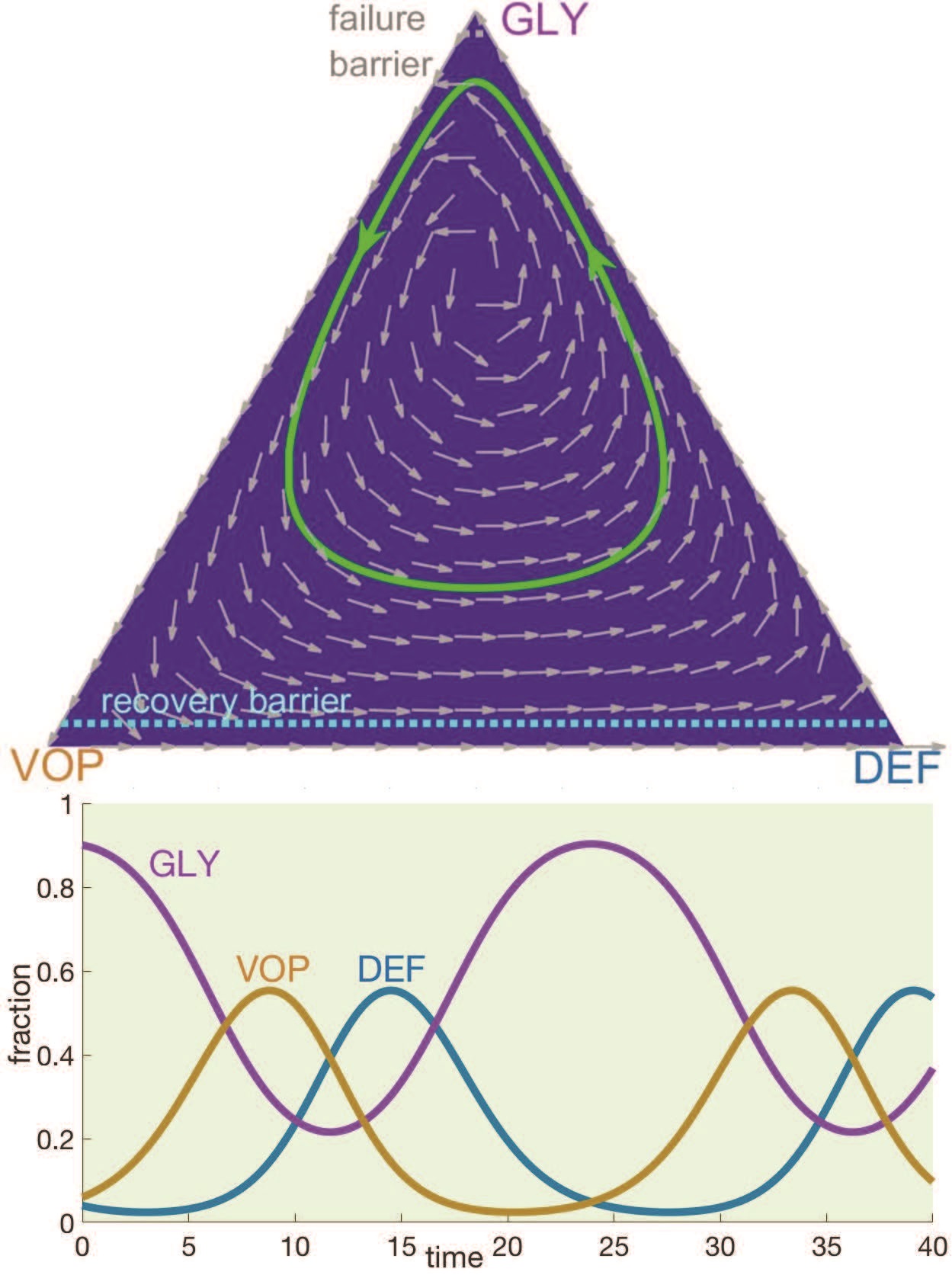}}%
\hspace{20pt}%
\subfigure[][]{%
\label{fig:ex4-b}%
\includegraphics[height=3in]{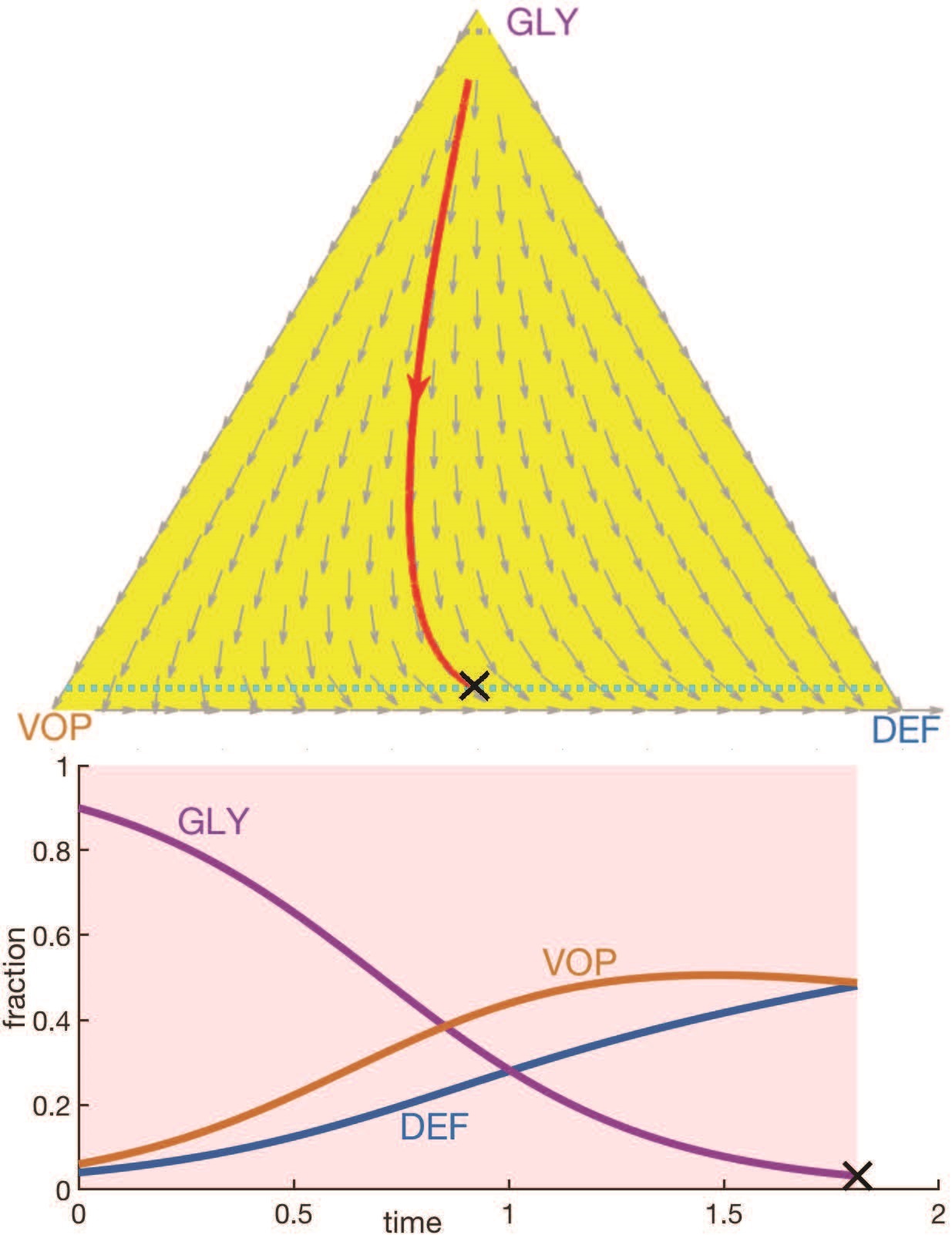}}

\caption[A set of two subfigures.]{\textbf{A comparison of two possible constant treatment scenarios} starting from an initial state $(\xD, \xG, \xV) = (0.04, 0.9, 0.06)$: \subref{fig:ex4-a} without any therapy; \subref{fig:ex4-b} with the MTD-based therapy.

\textbf{Top row:} phase portraits of corresponding vector fields (shown by \textit{gray arrows}) on a GLY-VOP-DEF triangle with illustrative trajectories. \textit{Blue background} and \textit{green reference trajectory}~-- no therapy at all. \textit{Yellow background} and \textit{red reference trajectory}~-- MTD-based therapy at all times.
\textit{Dash light blue} and \textit{gray lines} separate the recovery zone (bottom) and the failure zone (top) respectively.
\textit{Black cross}~-- termination due to crossing the recovery barrier.

\textbf{Bottom row:}
evolution of sub-populations with respect to time based on the reference trajectories above. \textit{Green time range}~-- no therapy. \textit{Pink time range}~-- MTD-based therapy.
\textit{Black cross}~-- termination due to crossing the failure or recovery barrier by GLY cells.
Note the different scaling of the time axis.

\textbf{Parameters:}
Following Figure 2 in \cite{Kaznatcheev2017a},   $\ba = 2.5$, $\bv=2$, $c=1$, $n=4$ and $\dmax=3$ for \subref{fig:ex4-b}.
The recovery and failure barriers are $\rb = \fb=10^{-1.5}$.
}
\label{fig:ex4}%
\end{figure}

For a range of parameter values (\ref{cyc}), this model predicts a heterogeneous regime\footnote{Other tumor regimes (fully angiogenic and glycolyctic) also exist outside of this parameter range.
They are less interesting from the point of view of treatment strategies, but we still consider them for the sake of completeness in Section \ref{other_regimes} of Supplementary Materials.} in cancer evolution with coexistent and oscillating proportions of GLY, VOP, and DEF.
Without any treatment, these sub-populations follow cyclic dynamics and a patient never recovers; see Figure \ref{fig:ex4-a}.

\begin{mybox}{Box 1: Mathematical model of the cancer sub-populations evolution}

\begin{minipage}{.4\textwidth}
       \textbf{Subpopulation Proportions}:

$(\xG, \, \xD, \, \xV)$

for GLY, DEF, and VOP 
respectively.\\

Note: $\xG+\xD+\xV=1.$
    \end{minipage}%
    \begin{minipage}{0.5\textwidth}
       \textbf{Relative Subpopulation Fitness:}  $\; (\psi^{}_G, \, \psi^{}_D, \, \psi^{}_V)$

       defined in Materials and Methods of Kaznatcheev et~al.\cite{Kaznatcheev2017a}).\\

       \textbf{Full Population Averaged Fitness:}

$<\psi> := \xG \psi^{}_G +\xD \psi^{}_D+\xV\psi^{}_V.$
    \end{minipage}
\bigskip

\textbf{Replicator equations to model the evolution of sub-populations:}
\begin{equation}\label{re}
\begin{cases}
\dot \xD = \xD(\psi^{}_D - <\psi>)
\\
\dot \xG = \xG(\psi^{}_G - <\psi>)
\\
\dot \xV = \xV(\psi^{}_V - <\psi>)
\end{cases}
\end{equation}

 Transformation/Reduction to a 2D system: \begin{equation}\label{trans}
\begin{cases}
q = \frac{\xV}{\xV+\xD}\\
p = \xG
\end{cases}
\text{or }~~
\begin{cases}
 \xD = (1-q)(1-p)\\
 \xG = p\\
 \xV = (1-p)q
 \end{cases}
\end{equation}

 \textbf{Equivalent dynamics of (\ref{re}) in reduced coordinates} (see Appendix B in \cite{Kaznatcheev2017a}):
\begin{equation}\label{re2d}
\begin{cases}
 \dot q = q(1-q)\Big(\frac{\bv}{n+1}\sum\limits_{k=0}^n p^k-c\Big)
\\
 \dot p = p(1-p)\Big(\frac{\ba}{n+1} - (\bv-c)q\Big)
 \end{cases}
\end{equation}

 \textbf{Parameters:}

 $\bullet \,$ $\ba$, the benefit per unit of acidification;

 $\bullet \,$ $\bv$, the benefit from the oxygen per unit of vascularization;

 $\bullet \,$ $c$, the cost of production VEGF;

 $\bullet \,$ $n$, the number of glycolytic (GLY) cells in the interaction group.

\bigskip

 \textbf{Conditions for homogeneous center equilibrium and periodic oscillations:}
 \begin{equation}\label{cyc}\frac{\ba}{n+1}<\bv-c < cn.\end{equation}

\bigskip

\textbf{Process \emph{terminates} as soon as either}
 \begin{equation}\label{barrier}
\begin{cases}
p(t) < \rb, &\quad\text{if therapy succeeds;}
\\
p(t) > 1- \fb,&\quad \text{if therapy fails.}
 \end{cases}
\end{equation}

\textbf{\emph{Terminal set}:}
\begin{equation}\label{ts}\Delta = \Big\{(q,p)\in [0,1]\times[0,1]: p < r^{}_ b \text{ or } p > 1 - \fb \Big\}.\end{equation}

\end{mybox}

Following Section 4.1. in Kaznatcheev et al. \cite{Kaznatcheev2017a} we consider a cell-type-targeting therapy that preferentially penalizes the fitness of GLY cells; see formula \eqref{rec} in Box~2.
During the treatment, a doctor defines the timing of therapy and its intensity.
This time-dependent intensity $d(t)$ can vary between $0$ (no therapy) and $\dmax>0$ (the maximum tolerated dose, MTD).
Two extreme case ($d(t) = 0$ versus $d(t) = \dmax$ for all times $t$) are illustrated in Figures \ref{fig:ex4-a} and \ref{fig:ex4-b} respectively.
In the latter case, GLY cells become extinct and the patient recovers quickly, however it is natural to ask whether this treatment strategy is optimal in some sense (i.e. could the recovery be much delayed if the patient received therapy less often or at a lower intensity?).
In the following section we will show that the MTD-based treatment can lead to an avoidably high cumulative amount of therapy (see Figure 2(c) or might even fail to achieve a recovery in situations where AT-based treatment would otherwise have succeeded (see Figure \ref{fig:ex6}), much like the early results from Zhang et al. in metastatic prostate cancer~\cite{Zhang2017a}.

\begin{mybox}{Box 2: Evolution dynamics with control on therapy intensity}
\textbf{Time-dependent intensity of GLY-targeting therapy:}
$\qquad d:\R_+\rightarrow [0,\dmax].$\\

\textbf{Evolutionary dynamics as a controlled system:}

\begin{equation}\label{rec}
\begin{cases}
 \dot q(t) = q(t)\Big(1-q(t)\Big)\Big(\frac{\bv}{n+1}\sum\limits_{k=0}^n p^k(t) -c\Big),
\\
 \dot p(t) = p(t)\Big(1-p(t)\Big)\Big(\frac{\ba}{n+1} - (\bv-c)q(t)-\boldsymbol{d(t)}\Big);
  \\
 q(0) = \qzero,~ p(0) = \pzero.
 \end{cases}
\end{equation}

\end{mybox}
One natural objective function to minimize is the total amount of therapy administered over the course of treatment (which in this case could be a surrogate for both toxicity and cost).
This can be quantified as $D = \int \limits_0^T d(t)dt,$ where the total time of treatment $T$ is dependent on the initial cancer subpopulation fractions and on our chosen therapy policy $d(\cdot)$.
However, this objective is problematic for two reasons.
First, the minimum of $D$ is clearly attained without any therapy (taking $d(t) \equiv 0$ implies $D = 0$ and $T=+\infty$), even though the dynamics become cyclic and the recovery is never achieved.
Second, if we constrain our minimization to only those $d(\cdot)$ that lead to recovery, an optimal treatment policy does not exist.
Instead, there is a sequence of treatment policies that lead to successively smaller $d$ values but with an unbounded increase in corresponding treatment times $T$.
The idea of such policies is simple: travel along the therapy-free trajectories of Figure \ref{fig:ex4-a} for most of the time, but use short bursts of therapy only when the drugs are most effective.
To approach the optimally small $d$, one would need to use shorter and shorter bursts, resulting in policies that are hard to implement in practice and would require unrealistically long treatment times $T$, but would yield a situation like a chronic disease, where while the tumor is never cured, it is always controlled.

In order to get a meaningful optimal policy we will penalize the treatment time by a \emph{time penalty} $\sigma>0$.
The total time spent on the treatment, including time when a patient skips the doses, is an important factor by itself.
Much longer time spent on a treatment causes worse quality of life and  additional costs for a patient.
Therefore, our objective is to minimize the sum of a \emph{therapy cost} $D$ and a \emph{treatment time cost} $\sigma T$, while guaranteeing the treatment results in recovery.
For every choice of $\sigma>0,$ the resulting treatment policies are thus {\em Pareto-optimal} with respect to $D$ and $T$.

In this paper we consider an \emph{objective function}
that is equal to $\int\limits_0^T d(t) \,dt \,+ \,\sigma T$, if a patient recovers, and is equal to $+\infty$ otherwise.
The {\em value function}, $u$, is defined as the minimum of this objective function (over the set of treatment policies), and any policy $d(\cdot)$ that realizes this minimum is called {\em optimal}.

Due to the structure of this optimization problem, one can show that virtually all optimal treatment policies are {\em bang-bang}: at any given time $t$, they either administer therapy at MTD-rate ($d(t)=\dmax$) or administer no drugs at all ($d(t)=0$);
see Section \ref{HJB_sec} in Supplementary Materials.
For such policies, the objective function 
becomes a weighted sum of the total {\em therapy time} $\widetilde T$ (when $d(t)\equiv \dmax$) and the total treatment time $T$, with $\dmax$ and $\sigma$ as the corresponding weights.
Moreover, this allows for a simple visual representation of any such policy: splitting the full state space into two parts (MTD dynamics vs. no-therapy dynamics), shown in yellow and blue respectively in all figures throughout this paper, and simplifies application of therapy into something familiar to clinicians: on or off.

\begin{mybox}{Box 3: Objective function}
\textbf{Terminal time (or the total treatment time):},
\begin{equation}\label{tt}T(\qzero,\pzero, d(\cdot)) = \min\Big\{t\in \R_+~|~ \boldsymbol (q(t),p(t))\in \Delta, ~ q(0) = \qzero,~ p(0) =\pzero\Big\}.\end{equation}

If the system never gets to the terminal set, we assume that $T(\qzero,\pzero, d(\cdot)) =+\infty.$

\textbf{Terminal cost function} is $g(q,p): \Delta \rightarrow \{0,+\infty \} $ s.t.

\begin{equation}\label{tc}
g(q,p) =
\begin{cases}
+\infty,&\quad \text{if } p > 1 - \fb,
\\
0,&\quad\text{otherwise.}
\end{cases}
\end{equation}

Let $  T := T( \qzero,\pzero, d(\cdot))$ be the terminal time.

 \textbf{Treatment cost (objective) function}:
\begin{equation}\label{obj}J(\qzero,\pzero ,  d(\cdot) )=\int\limits_0^{T} (d(s)+ \sigma)ds+g\Big(q\big( T\big), p\big( T\big)\Big).\end{equation}

 $J$ is finite if the system (\ref{rec}) terminates at the recovery barrier, and is infinite otherwise (i.e., if the system terminates at the failure barrier or does not terminate at all).

\textbf{Value function}:
\begin{equation}\label{vf}u(\qzero,\pzero) = \inf\limits_{d(\cdot)} J(\qzero,\pzero,  d(\cdot) )\end{equation}
can be found by solving Hamilton-Jacobi-Bellman {\bf (HJB)} PDE:

\begin{equation}\label{IHJB}
\min\limits_{d \in [0, \dmax]}
\left\{\nabla u(q, p)\cdot \left(
                                                                      \begin{array}{c}
                                                                        \dot q (q,p,d)\\
                                                                        \dot p (q,p,d)\\
                                                                      \end{array}
                                                                    \right)+ d + \sigma\right\} = 0,
                                                                    \qquad (q,p)\in ([0,1]\times[0,1])/\Delta.
\end{equation}

\textbf{The boundary conditions} of HJB equation:

\begin{equation}\label{bc}
\begin{cases}
u(q,p) = 0,&\quad \text{ if } p < \rb;\\
u(q,p) = +\infty, &\quad \text{ if } p
>1-\fb.
\end{cases}
\end{equation}

Once $u$ and its gradient are found through a numerical approximation, they can be used to obtain the optimal control {\em in feedback form:} $d^* = d(q, p).$

\end{mybox}

\section{Results}\label{results}

\subsection{Quantifying the benefits of optimal treatment strategies }\label{Quantifying_the_benefits}
In this section we
apply the optimal control theory to an example considered in Kaznatcheev~et~al.\cite{Kaznatcheev2017a}. The details of the optimal control problem formulation are summarized in Box 3.
We take the same model parameters as in Figure 2  of Kaznatcheev~et~al.\cite{Kaznatcheev2017a}:
 $\ba = 2.5$, $\bv=2$, $c=1$, $n=4,$ strength of MTD $\dmax = 3$, and initial state $(\qzero,\pzero) = (0.6, 0.9)$, which by formula \eqref{trans} is equivalent to $(\xD, \xG, \xV) = (0.04, 0.9, 0.06)$.   We also use $\sigma=0.01$ to incorporate a time-penalty absent in the original model. We take $\rb=\fb = 10^{-1.5}$ as recovery
 and failure barriers\footnote{This change in parameter values is meant to decrease the computational cost of our numerical approach (see section \ref{TD} in Supplementary Materials.)  The original $\rb=\fb = 10^{-4}$ from Kaznatcheev~et~al.\cite{Kaznatcheev2017a} would require computations on a finer mesh.}

\begin{figure}[H]
\centering%

\settoheight{\tempdima}{\includegraphics[width=.8\linewidth]{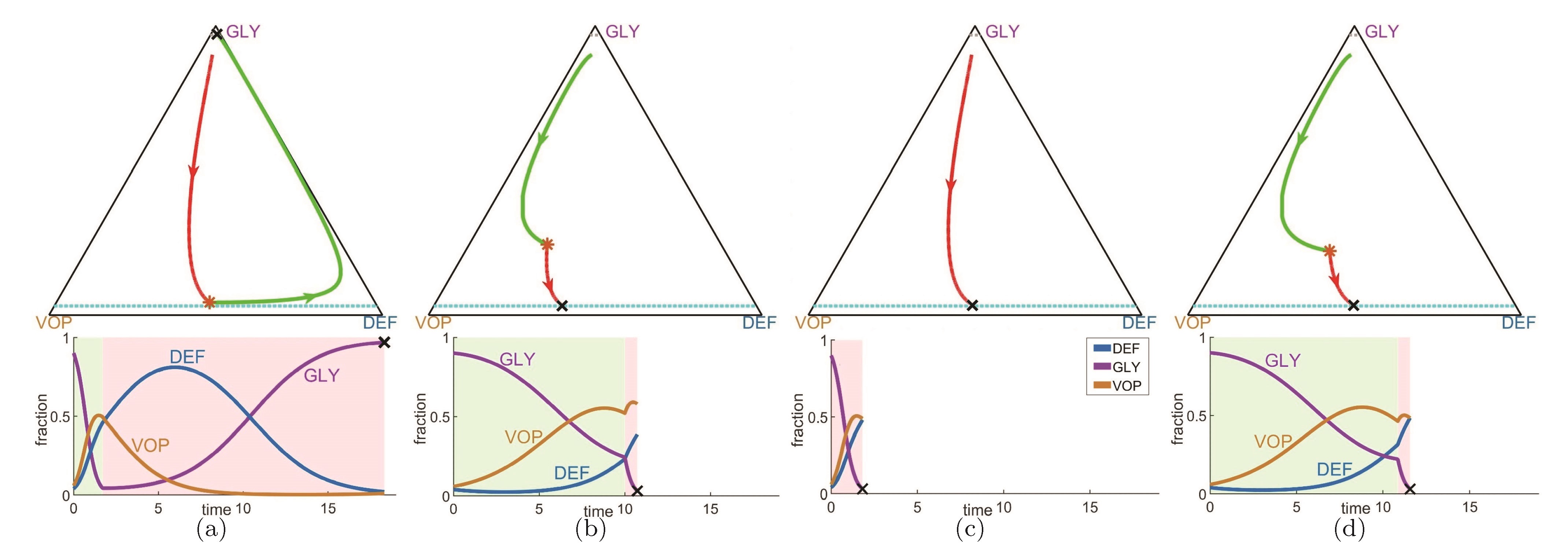}}%
\centering\begin{tabular}{@{}c@{ }c@{}}

\includegraphics[width=1.0\linewidth]{Fig2.jpg}
\end{tabular}

{\small
\centering\renewcommand{\arraystretch}{1.2}
\begin{tabular}{
!{\vrule width 2pt}c|c|c|c|c!{\vrule width 2pt}}
\ChangeRT{2pt}
Subfigure& Policy & Total time & Time till switching (denoted by *) & Overall cost \\
\ChangeRT{2pt}
(a) & ``treat immediately but not long enough''
& 18.32 & 1.7 & $+\infty$ \\
\hline
(b)&``start treating after 10 time steps''
& 10.73 & 10 & 2.29\\
\hline
(c)& ``MTD-based''
& 1.81 & --- & 5.45\\
\hline
(d)& ``optimal''
& 11.56 & 10.86 & 2.25\\
\ChangeRT{2pt}
\end{tabular}
}
\caption[A set of four subfigures.]
{\textbf{The importance of optimal scheduling for drug therapy: ensuring recovery and decreasing the cost of treatment.}

Tumor evolution under four different treatment strategies for the same initial state $(\xD, \xG, \xV) = (0.04, 0.9, 0.06)$.
A seemingly reasonable treatment strategy may not lead to a recovery; see subfigure  (d).
Even if a patient eventually recovers (as in subfigures (b) and (c), the overall cost of treatment can be reduced by pursuing a provably optimal policy (d).

\textbf{Top row:} tumor evolutionary trajectories under different strategies on a GLY-VOP-DEF triangle.
\textit{Green part of a trajectory} -- no therapy is used.
\textit{Red part of a trajectory} -- MTD-based (standard) therapy.
The moment of switching from one regime to another is denoted by (red *).
\textit{Dash light blue} and \textit{gray lines} separate the recovery zone and the failure zone respectively.
\textit{Black cross} -- termination due to crossing the failure or recovery barrier.

\textbf{Bottom row:}
evolution of sub-populations with respect to time based on the reference trajectories above.
\textit{Green time range} -- no therapy.
\textit{Pink time range} --  MTD-based therapy.
\textit{Black cross} -- termination due to crossing the failure or recovery barrier by GLY cells.

\textbf{Parameters:}
Following Figure 2 in \cite{Kaznatcheev2017a} and Figure \ref{fig:ex4}, we use
the initial state $(\xD, \xG, \xV) = (0.04, 0.9, 0.06)$; game parameters: $\ba = 2.5$, $\bv=2$, $c=1$, $n=4$; the MTD rate: $\dmax=3$; the recovery and failure barriers: $\rb = \fb=10^{-1.5}$; and the time-penalty: $\sigma = 0.01$.
}
\label{fig:ex1}
\end{figure}

In Figure \ref{fig:ex1} we compare the treatment cost (\ref{obj}) and treatment time (\ref{tt}) of trajectories corresponding to four different treatment strategies.

The first two strategies we consider are similar to those modeled in Kaznatcheev~et~al.\cite{Kaznatcheev2017a}: 2(a) is an example of a bad policy that may cause a failure by stopping the therapy prematurely, while 2(b) is a good policy based on ad-hoc adjustment of the start time for the therapy.
We also illustrate a ``MTD-based''' policy 2(c), which is analogous to the standard of care using MTD as long as possible.
Even though both 2(b) and 2(c) lead to recovery, neither of these is optimal (with the MTD-based approach resulting in excessive amount of drugs, captured by the higher cost).
The policy minimizing our objective function can be found by solving the HJB equation; we illustrate it in 2(d).

\begin{figure}[H]
\centering

\subfigure[][]{%
\label{fig:ex0-a}%
\includegraphics[height=1.8in]{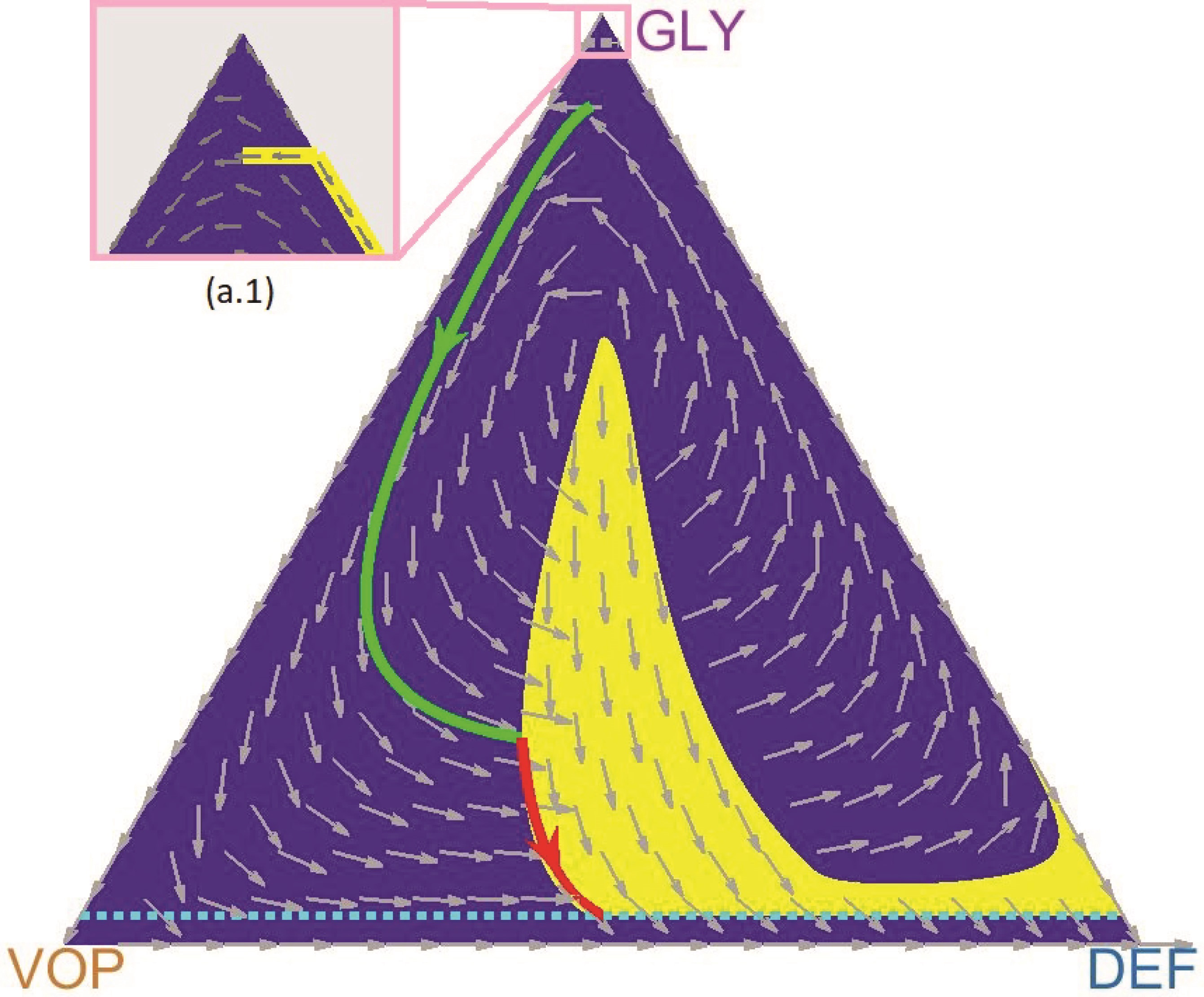}}%
\hspace{8pt}%
\subfigure[][]{%
\label{fig:ex0-b}%
\includegraphics[height=1.8in]{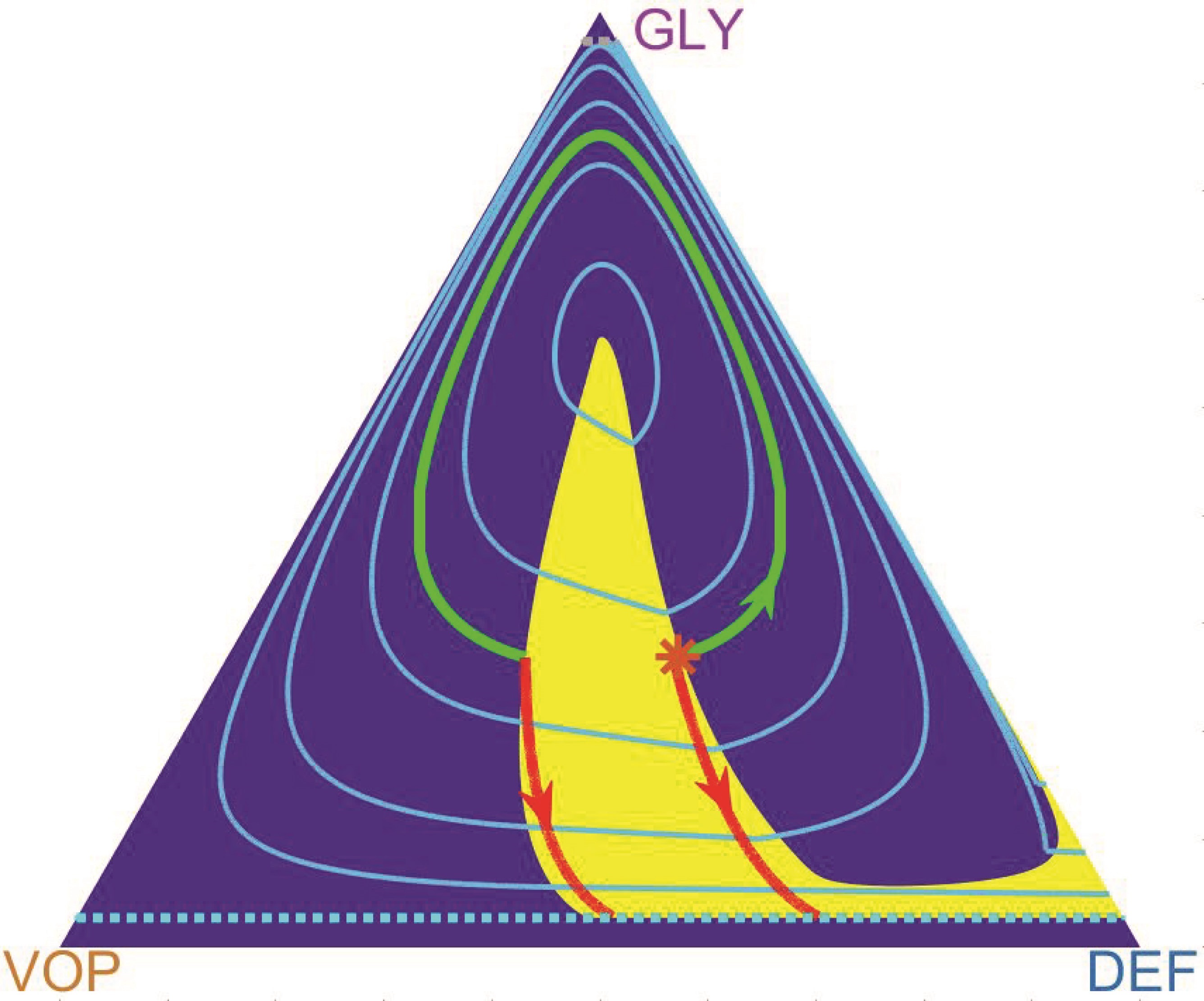}}
\subfigure[][]{%
\label{fig:ex0-c}%
\includegraphics[height=1.8in]{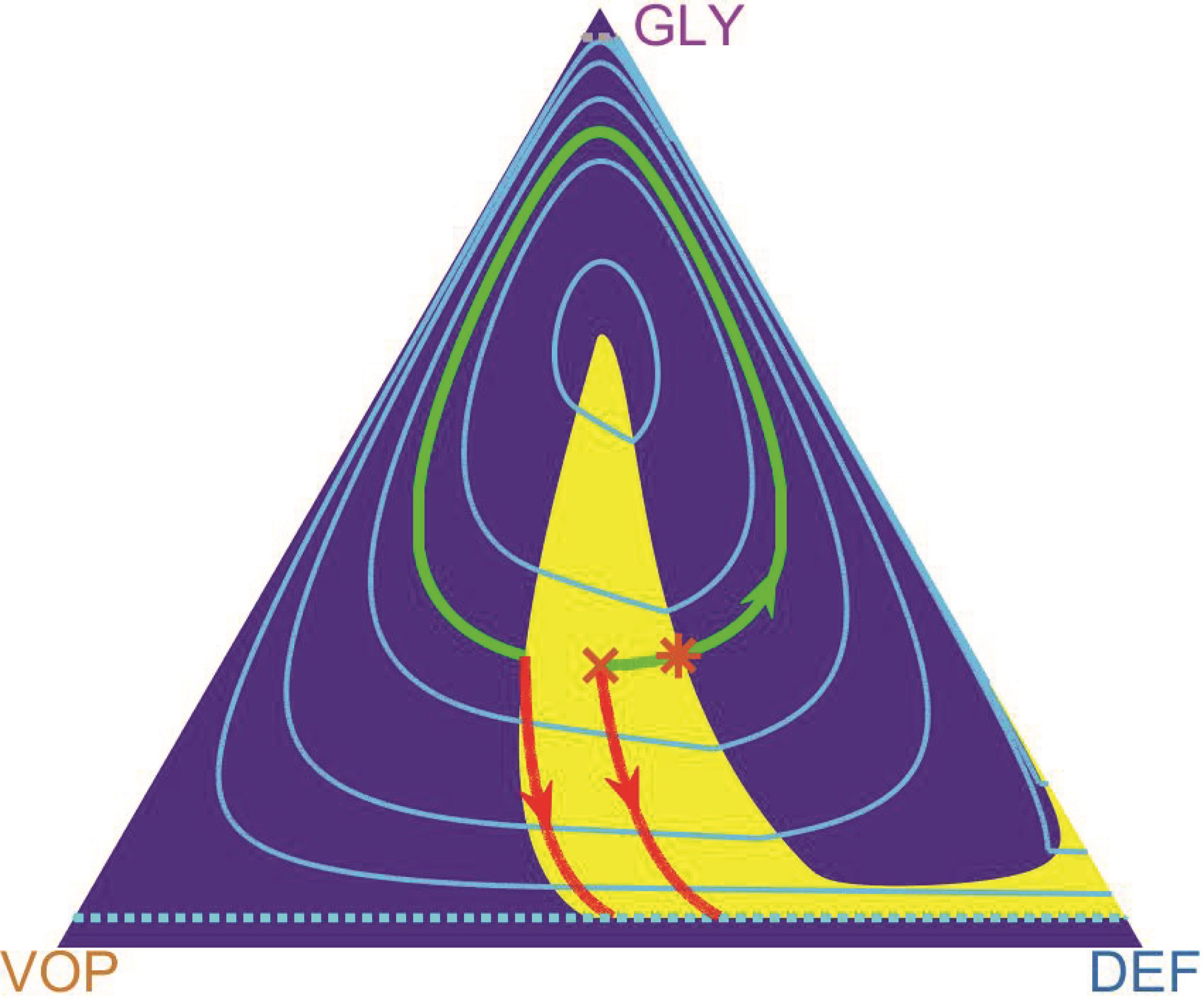}}
\caption[]{\textbf{Optimal control in feedback form, the value function, and the pitfalls of PMP.}

\textbf{\subref{fig:ex0-a}}:  A phase portrait of the optimal system dynamics.
The vector field is shown by \textit{gray arrows} over the
optimal drugs-off (\textit{blue background}) and drugs-on (\textit{yellow background}) regions.  A sample optimal trajectory (in green and red) corresponds to the initial state from Figure \ref{fig:ex1}.

\textbf{\subref{fig:ex0-b}}:  Computation of the value function $u$ (whose level curves are shown by \textit{light blue lines}) is used to determine the optimal drugs-on and drugs-off regions (shown in yellow and dark blue respectively).
Optimal trajectories are not unique for initial states on the shockline (where the level curves of $u$ are not smooth).  Two such optimal trajectories are shown starting from an asterisk (*).

The green-red trajectory takes longer to reach the recovery, but uses less drugs than the red (start-drugs-right-away) trajectory.  The cumulative cost is the same for both of them.

\textbf{\subref{fig:ex0-c}}: For initial conditions off the shocklines of $u$, there can still be multiple {\em locally optimal} trajectories.  We show an example of two such trajectories starting from a cross marker ($\times$). The risk of applying the PMP method is that it might yield either of them, but only the red (start-drugs-right-away) is globally optimal.

\textbf{Parameters:}
$\ba = 2.5$, $\bv=2$, $c=1$, $n=4$, $\dmax=3$, $\sigma= 0.01$, $\rb = \fb=10^{-1.5}$. Initial states of trajectories: \subref{fig:ex0-a}  $(\xD, \xG, \xV) = (0.04, 0.9, 0.06)$;  \subref{fig:ex0-b} denoted by (*)  $(\xD, \xG, \xV) = (0.417, 0.311, 0.272)$;  \subref{fig:ex0-c} denoted by $(\times)$   $(\xD, \xG, \xV) = (0.35, 0.3, 0.35).$
}
\label{fig:ex0}%
\end{figure}

The corresponding therapy on/off regions and the resulting vector field are shown in Figure \ref{fig:ex0-a}.
The zoomed version shows that trajectories can be prevented from crossing the failure barrier by using the MTDs just before crossing.
In fact, a {\em chattering control} (with intermittent and sufficiently frequent use of MTDs) would be sufficient to guarantee this.

The level sets of
$u$ in Figure \ref{fig:ex0-b} show that value functions need not be smooth.
Since the gradient of $u$ is used to determine the optimal course of action (therapy on or off), there can actually be more than one optimal policy for initial states on a \textit{shockline} (where that gradient is undefined).
We show an example of such trajectories for an initial point $(\xD, \xG, \xV) \approx (0.417, 0.311, 0.272)$ (solid green and red lines in Figure \ref{fig:ex0-b}).
Both trajectories yield the same cost of $2.764$.
Moreover, non-smoothness of the value function often poses a challenge for method's based on Pontryagin Maximum Principle (PMP) \cite{Pontryagin1962}
even if the initial state {\em is not} on a shockline. For example, perturbing the initial state to a nearby $(\xD, \xG, \xV) = (0.35, 0.3, 0.35)$, denoted by a cross in Figure \ref{fig:ex0-c}, one sees two {\em locally optimal} trajectories and PMP might yield either of these depending on the initial guess.
The green one is however inferior to the {\em globally} optimal red trajectory, which is always recovered by solving the HJB equation.

\subsection{Optimization trade-offs: total administered drugs versus time to recovery}

\begin{figure}[H]
\centering%

\includegraphics[width=1.0\linewidth]{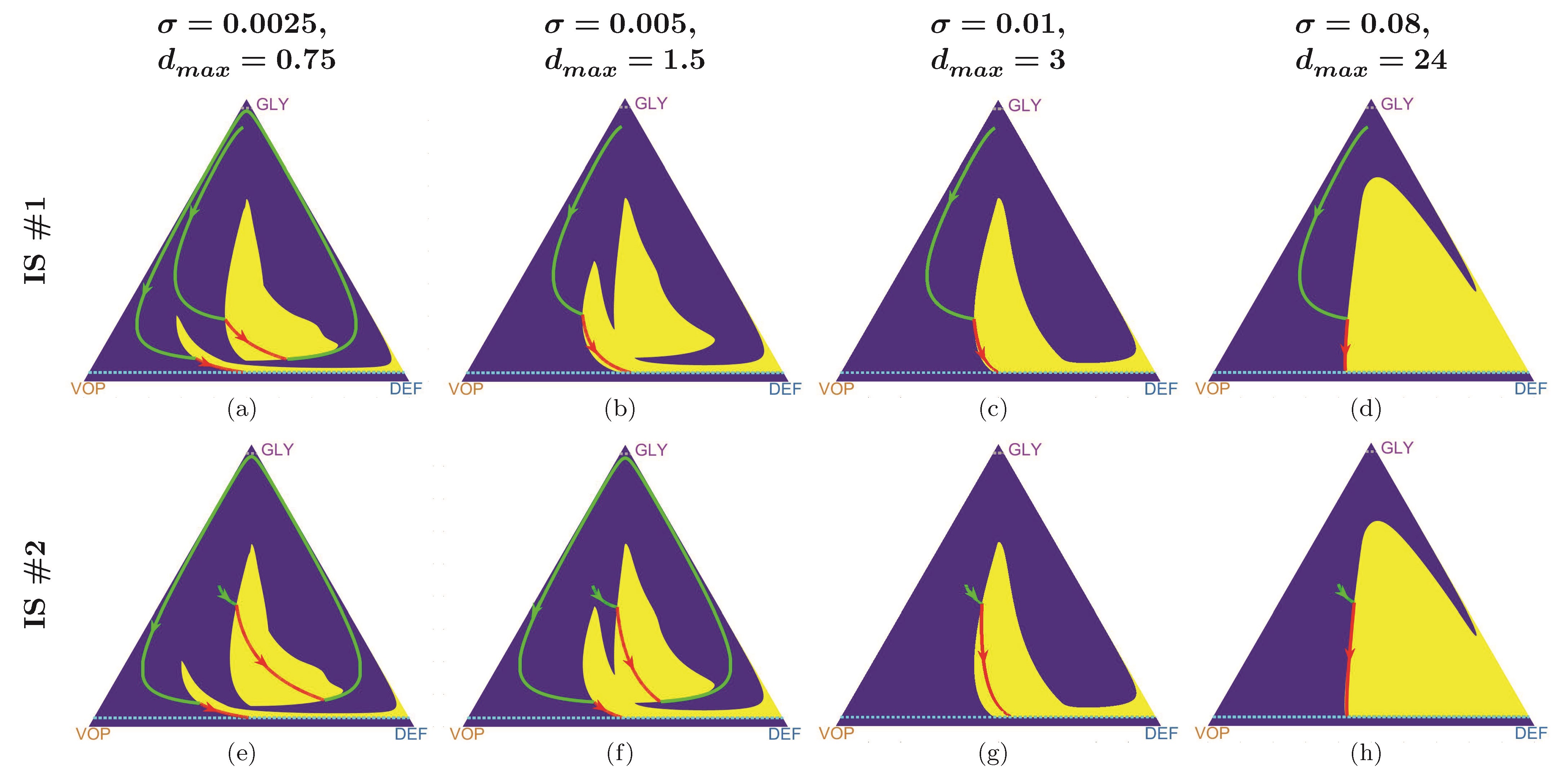}

\bigskip

{\small
\centering
\begin{tabular}{
!{\vrule width 2pt}c|c!{\vrule width 2pt}
c|c|c|c!{\vrule width 2pt}
c|c|c|c!{\vrule width 2pt}}
\ChangeRT{2pt}
\multicolumn{2}{!{\vrule width 2pt}c!{\vrule width 2pt}}{Parameters} & \multicolumn{4}{c!{\vrule width 2pt}}{$\begin{array}{c}\text{Initial State (IS) }\#1:\\ (\xD, \xG, \xV) = (0.04, 0.9, 0.06)\end{array}$  } & %
    \multicolumn{4}{c!{\vrule width 2pt}}{$\begin{array}{c}\text{Initial State }\#2:\\ (\xD, \xG, \xV) = (0.15, 0.5, 0.35)\end{array}$}\\
\ChangeRT{2pt}
$\sigma$ & $\dmax$ &
subfigure & total time & total drugs & overall cost &
subfigure & total time & total drugs & overall cost \\
\ChangeRT{2pt}
 0.0025 & 0.75 &
 (a) & 42.7519 & 2.0808 & 2.1877 &
 (e) & 33.9242 & 3.0750 & 3.1598 \\
\hline
 0.005 &  1.5 &
 (b) & 11.5998 & 2.1376 & 2.1956 &
 (f) & 33.1649 & 3.0254 & 3.1912 \\
\hline
 0.01 &  3 &
 (c) & 11.5649 & 2.1306 & 2.2463 &
 (g) & 2.8230 & 3.1241 & 3.1524 \\
\hline
 0.08 &  24 &
 (d) & 10.9820 & 2.1520 & 3.0306 &
 (h) & 1.8920 & 3.1490 & 3.3004\\
\ChangeRT{2pt}
\end{tabular}
}
\caption[]{\textbf{Varying the MTD level affects all optimal trajectories. }

Here we illustrate a fixed $\sigma / \dmax$ ratio (with $\dmax$ increasing from left to right), which is equivalent to preserving the relevant importance (trade-off) between the total therapy time $\tilde T$ and the total treatment time $T.$
Nevertheless, the optimal drugs-on regions (in yellow) vary since any changes in $\dmax$ also affect the dynamics of the system (\ref{rec}).

\textbf{Parameters:}
$\ba = 2.5$, $\bv=2$, $c=1$, $n=4$; $\rb = \fb=10^{-1.5}$. Two initial states and the $(\sigma, \dmax)$ values are specified in the  table above.

}
\label{fig:ex5}%
\end{figure}

The optimal therapy-on regions are clearly dependent on  specific values of all model parameters.  Here we explore their dependence on $\sigma$ and $\dmax$.
Recall that, for every policy leading to recovery, the {\em overall cost} of treatment is a sum of the  ``therapy cost'' (i.e., the total amount of drugs administered, $D = \int\limits_0^T d(t) dt$) and the treatment-time cost $\sigma T$.
Since the optimal control is bang-bang, this can be re-written as a weighted sum of the time-till-recovery $T$, and the total drug therapy time $\tilde{T} \leq T$.
That is, for controls based on repeated therapy-off/MTD-level-therapy switches, we can re-write the overall cost as
$J = \dmax \tilde{T} + \sigma T,$ with the ratio between the weights $(\sigma/\dmax)$ representing the ``relative importance'' of $T$ and $\tilde{T}$ for the optimization.
But the functional role of these weights is quite different: while $\sigma$ can be chosen to reflect our preferences, the MTD-rate $\dmax$ is dictated by the medical reality, which will be patient and drug specific.
By varying $\dmax$ while keeping $(\sigma/\dmax)$ constant, we can study the role played by the MTD-level in determining optimal policies under a fixed relative preference between the objectives.
In Figure \ref{fig:ex5} we conduct this experiment for two different initial states and the same set of game parameters  ($\ba = 2.5$, $\bv=2$, $c=1$, $n=4$), demonstrating that the value of $\dmax$ strongly influences the optimal policies and the shape of the ``therapy-on'' yellow regions.

\begin{figure}[H]
\centering%

\includegraphics[width=0.8\linewidth]{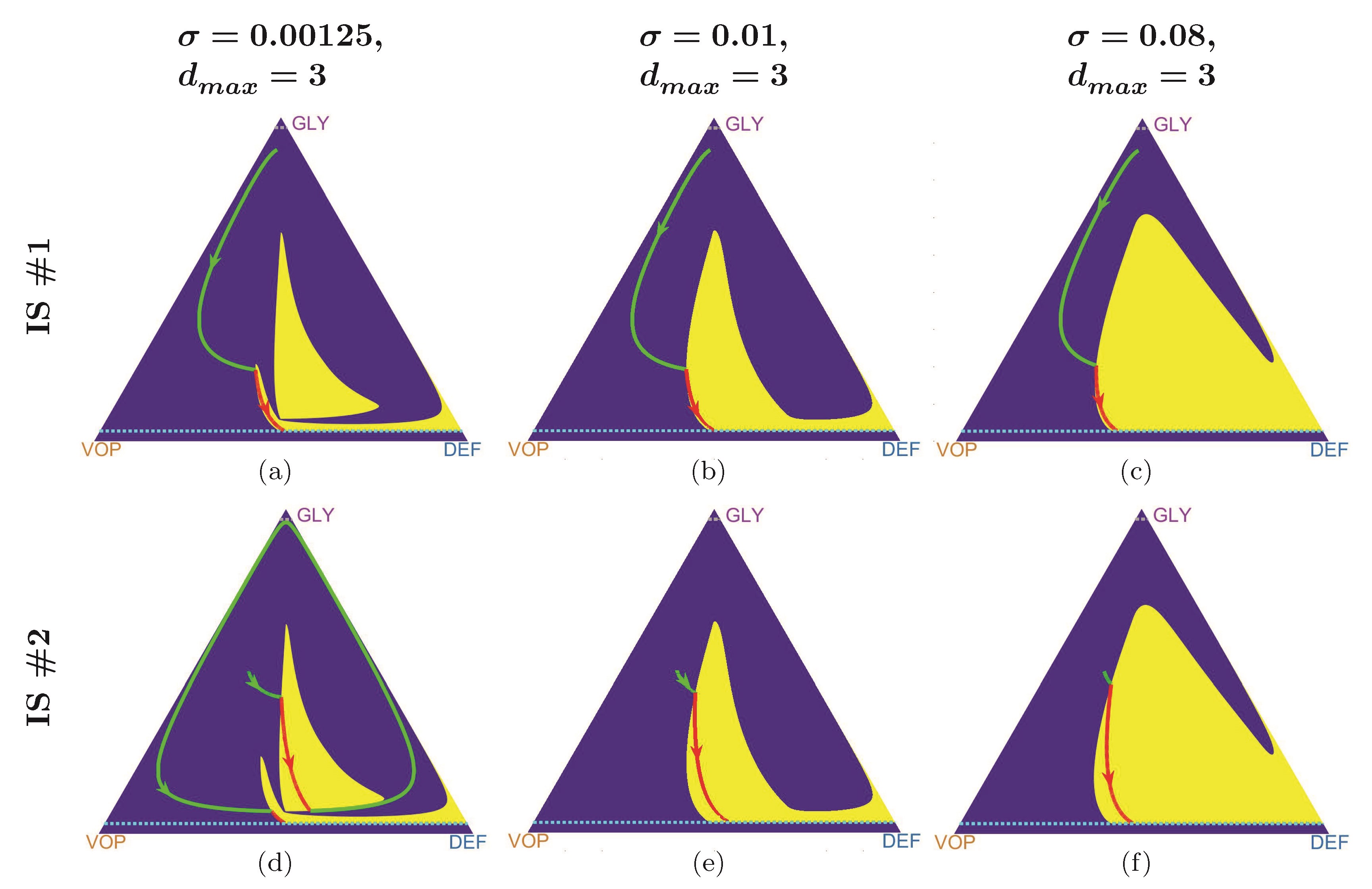}

\bigskip

{\small
\centering
\begin{tabular}{
!{\vrule width 2pt}c|c!{\vrule width 2pt}
c|c|c|c!{\vrule width 2pt}
c|c|c|c!{\vrule width 2pt}}
\ChangeRT{2pt}
\multicolumn{2}{!{\vrule width 2pt}c!{\vrule width 2pt}}{Parameters} & \multicolumn{4}{c!{\vrule width 2pt}}{$\begin{array}{c}\text{Initial State (IS) }\#1:\\ (\xD, \xG, \xV) = (0.04, 0.9, 0.06)\end{array}$  } & %
    \multicolumn{4}{c!{\vrule width 2pt}}{$\begin{array}{c}\text{Initial State }\#2:\\ (\xD, \xG, \xV) = (0.15, 0.5, 0.35)\end{array}$}\\
\ChangeRT{2pt}
$\sigma$ & $\dmax$ &
subfigure & total time & total drugs & overall cost &
subfigure & total time & total drugs & overall cost \\
\ChangeRT{2pt}
 $0.00125$ & $3$ &
 (a) & 11.6363 & 2.1004 & 2.1447 &
 (d) & 34.4945 & 3.0713 & 3.1144 \\
\hline
 $0.01$ & $3$ &
 (b) & 11.5649 & 2.1306  & 2.2463&
 (e) & 2.8230 & 3.1241  & 3.1524\\
\hline
 $0.08$ &$ 3$ &
 (c) & 10.9897 & 2.1574 & 3.0366 &
 (f) & 1.9470 & 3.1648 & 3.3206 \\
\ChangeRT{2pt}
\end{tabular}
}
\caption[]{\textbf{Different trade-offs (time to recovery vs total drugs) yield different optimal trajectories.}

The MTD-rate $\dmax$ is fixed while $\sigma$ increasing from left to right. The ratio $(\sigma / \dmax)$ defines the relevant importance of the total amount of drugs (therapy cost) $d$ versus  the total treatment time $T$. An increase in $\sigma$ results in smaller $T$ and larger $d$ along the optimal trajectories.

\textbf{Parameters:} $\ba = 2.5$, $\bv=2$, $c=1$, $n=4$; $\rb = \fb=10^{-1.5}$.}
\label{fig:ex11}%
\end{figure}

\begin{figure}[H]
\centering

\includegraphics[width=0.9\linewidth]{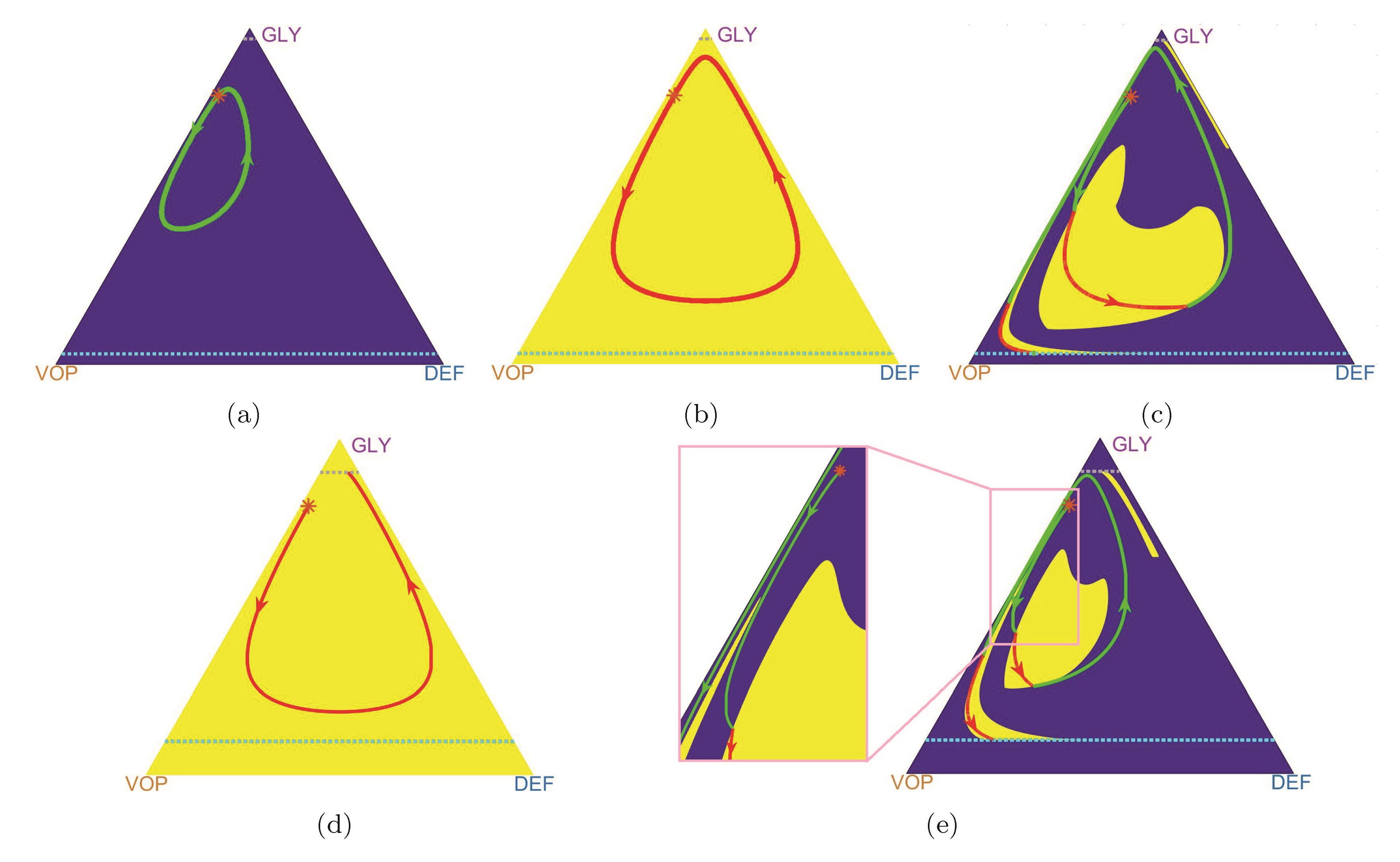}
\caption[]{
\textbf{MTD-based policy versus the optimal (adaptive) policy when the MTD rate is low (\boldsymbol{$\dmax< \frac{\ba}{n+1}$}).}

\textbf{Top row:} trajectories under both (a) ``no therapy'' policy and (b) the MTD-based policy are cyclic and cannot cross either the recovery nor the failure barrier from the initial state denoted by (*).
Nevertheless, the adaptive/optimal switching leads to a full recovery (c).

\textbf{Bottom row:} With a larger failure zone, an MTD-based policy leads to patient's putative death (d) even though it is still possible to cross the recovery barrier under an adaptive/optimal policy (e).

\textbf{Parameters:} $\ba = 4$, $\bv=2$, $c=1$, $n=4$; $\dmax=0.3$, $\sigma = 0.03$ and the initial state
(*) $(\xD, \xG, \xV) = (0.02, 0.8, 0.18).$ Top row  $\rb = \fb = 10^{-1.5}$, bottom row  $\rb = \fb = 10^{-1}.$
 }
\label{fig:ex6}%
\end{figure}

On the other hand, for any fixed/biological $\dmax$ value, we can vary $\sigma$ to study how the trade-off between $\tilde{T}$ and $T$ affects the optimization.
This experiment is conducted for the same two initial states in Figure \ref{fig:ex11}.
As we can see, smaller $\sigma$ entails larger total time.
Intuitively, this happens since it becomes ``cheaper'' to pause the therapy until we reach a ``better'' state to administer the drugs.
Larger $\sigma$ leads to a shorter time-to-recovery $T$, but also an increase in the total amount of administered drugs $D = \dmax \tilde{T}$ and a larger therapy-on region (shown in yellow) in the state space.

\subsection{``Incurable'' states and periodic trajectories under MTD treatment.}\label{sec_ha}

One might think that, despite being sub-optimal, an aggressive MTD-based strategy is at least always fully reliable and
the resulting trajectories are guaranteed to reach the recovery zone from every initial configuration, as shown in Figure 2(c).  Indeed,  if $\frac{\ba}{n+1} \leq \dmax$, the MDT-based policy ($d(t) \equiv \dmax $) guarantees that $\dot{p}$ is always negative; see equation~\eqref{rec}.  But with $\frac{\ba}{n+1}> \dmax$ the recovery might not be attained with the constant use of MTDs  (even if some other treatment policies are successful!).

Consider, for example, the following set of  parameters: $\ba = 4$, $\bv=2$, $c=1$, $n=4$; $\rb = \fb = 10^{-1.5}$; $\dmax=0.3$, $\sigma = 0.03$ and an initial state $(\xD, \xG, \xV) = (0.02, 0.8, 0.18).$
Under these parameters the  MTD-based therapy has a periodic trajectory\footnote{This is easy to prove by redefining the parameter
$\ba := \ba - (n+1)\dmax>0$
and reducing the MTD-based case to the periodic behavior of the original uncontrolled system \eqref{re2d} in the parameter regime \eqref{cyc}.}; see Figure 6(b).
Since the treatment time is infinite, the cost (\ref{obj}) of such a policy is $+\infty$.
(In reality this would lead to the emergence of drug resistance and eventual failure, but this biological situation is not modeled in Kaznatcheev~et~al.\cite{Kaznatcheev2017a}.)

We can see that neither of two extreme strategies (``no-drugs-at-all'' in Figure 6(a) and the MTD-based ``drugs-all-the-time'' in Figure 6(b)) can bring the trajectory to the recovery zone.
However, their adaptive combination can still achieve the objective.   We show a trajectory corresponding to the optimal policy in Figure 6(c).  With a larger failure zone (e.g., $\rb = \fb = 10^{-1}$), a previously successful MTD-based treatment might even result in death (Figure 6(d)), while the adaptive strategy still leads to recovery (Figure 6(e)).

\begin{figure}[H]
\centering%

\includegraphics[width=.6\linewidth]{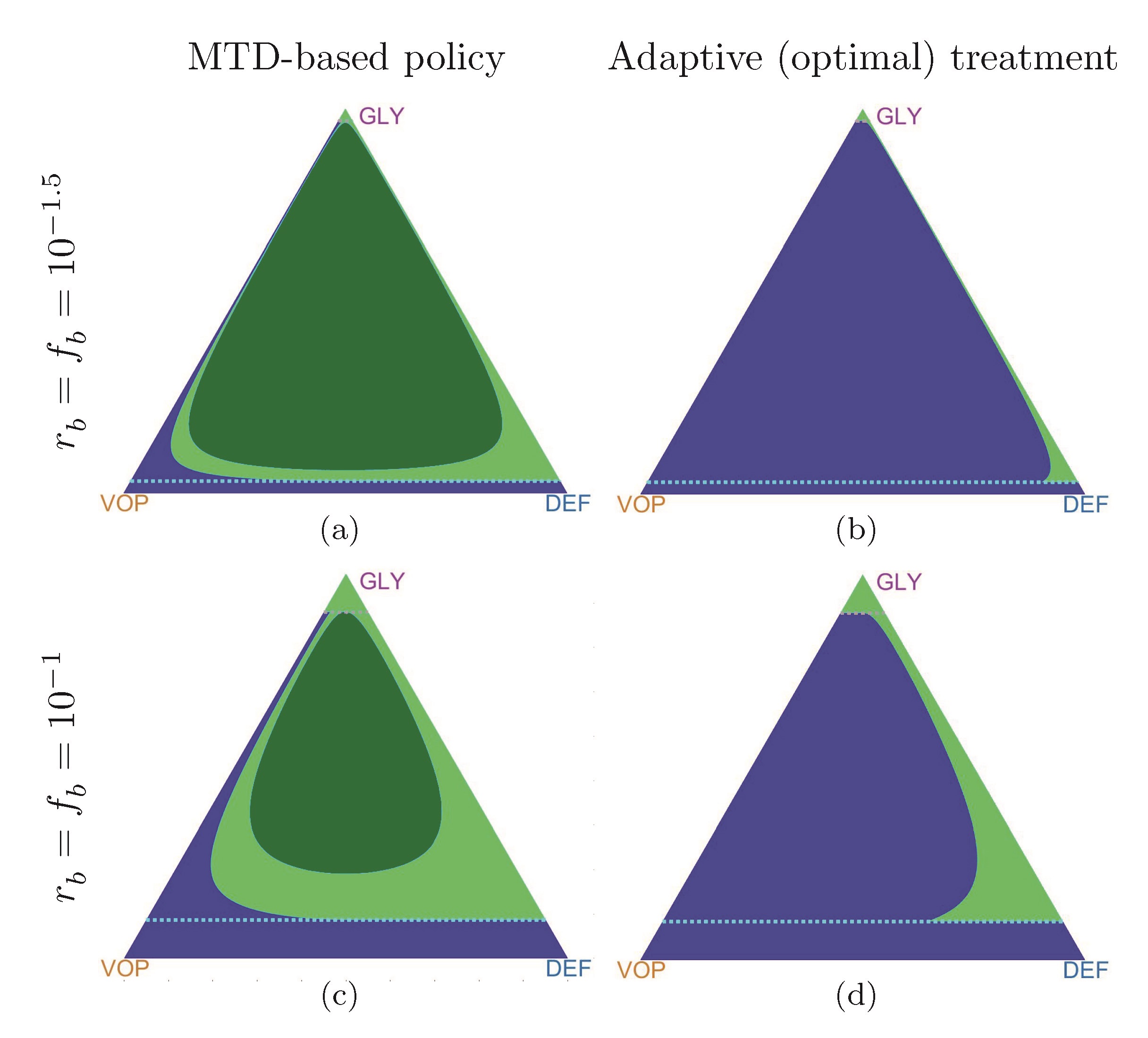}

\caption[]{\textbf{Comparison of the ``incurable'' area for the MTD-based policy versus the adaptive policy.}

Using adaptive strategies even with small MTD the patient can recover from most of the states that is impossible using MTD-based policy.
\textit{The blue color} indicates initial states from which the corresponding trajectories are able to achieve the recovery zone; \textit{the green color}  represents states from which recovery is impossible (``incurable'' area): \textit{the light green color} --  trajectories eventually get into the failure zone; \textit{the dark green color} --  trajectories are cyclic and cannot cross either the recovery or failure zones.

\textbf{Parameters:}
$\ba = 4$, $\bv=2$, $c=1$, $n=4$; $\dmax = 0.3$.
}
\label{fig:ex7}%
\end{figure}

For a fixed treatment policy, we define its corresponding \emph{``incurable'' area} to be a set of states starting from which it is impossible to cross the recovery barrier. For example in Figure 6, the incurable area of the MTD-based policy includes the state $(\xD,\xG,\xV) = (0.02,0.8,0.18)$  (when $\rb = \fb = 10^{-1.5}$ or $\rb = \fb = 10^{-1}$).
However, this state is \emph{not} in the (dramatically smaller) incurable area of the adaptive/optimal policy; see Figure \ref{fig:ex7}.

Starting from any incurable configuration, one could similarly pose a different control problem of {\em maximizing} the time until crossing the failure barrier.  While we do not address it here, we note that the HJB approach would be quite suitable to find optimal treatment policies for this problem as well.

Of course, the ``incurable areas'' are highly dependent on model parameters.
In section \ref{ss:incurable_param_var} of Supplementary Materials, we show that they can grow due to an increase in the MTD rate $\dmax$ or a decrease in vascularization benefits $\bv$.

\section{Discussion}

By now it is widely accepted that cancer is an evolutionary process, and that variation and selection drive the emergence of drug resistance.
While this new knowledge is driving cancer research forward, it has largely not yet affected clinical practice, with the majority of clinical protocols relying on MTD-based approaches, which invariably fail in the setting of most metastatic disease.
This is changing with the advent of adaptive therapy -- therapeutic strategies specifically designed with a changing regimen prescribed: one that adapts to an evolving tumor~\cite{Stankova2018a}.

To optimally design AT protocols,
the underling dynamics of the tumor growth and treatment response must be known.
While the methods of learning these dynamics are still in their infancy, the last decade has seen a flurry of activity using evolutionary game theory and other evolutionary models
for a range of prescribed dynamics.
These works have shown qualitative changes in tumor behavior in a range of treatment scenarios, and importantly,
demonstrated that
the order~\cite{Nichol2015}, sequence~\cite{Kaznatcheev2017a} and timing~\cite{Basanta2012} of therapy can drastically change the outcomes.
As we come closer to the reality of evolutionarily designed therapeutic trials in the mainstream, it is important then that we develop methods not just to make our outcomes better, but also to formally optimize them.

However, before using the mathematical tools for optimization, one also needs to choose a specific quantifiable criterion for comparing the outcomes.
Once that criterion is selected and the underlying mathematical model is sufficiently accurate, the best treatment strategy can be found by the techniques of optimal control theory.
In this paper, we show how this can be done for one particular heterogeneous cancer model previously described in Kaznatcheev~et~al.\cite{Kaznatcheev2017a}.
Given a set of model parameters ($\bv, ~\ba, ~c, ~n$) and treatment/recovery parameters ($\dmax, \rb, \fb$), we can find an optimal treatment policy for any initial distribution of cells ($\qzero,~ \pzero$).
We show that the optimal treatment policy can have multiple regimes: always on $ d^*(\cdot)\equiv \dmax$, always off $ d^*(\cdot)\equiv0$, and involving several contiguous treatment periods.
For the latter, the challenge is to accurately approximate the on/off ``switching curves'' in the state space.
We show that the definition of optimal treatment policy is heavily dependent on a parameter $\sigma$, a ``cost'' term, describing the relative importance of minimizing the total amount of drugs administered versus the total time to recovery.
We further show that, for some parameter regimes, there are ``incurable regions'' in the state space -- the starting configurations that will not lead to a recovery regardless of the chosen treatment strategy -- suggesting an alternative therapy (or goal) should be considered.
Moreover, for some other starting configurations, the ``always on'' strategy might not lead to a recovery even if the recovery is actually achievable with some on/off hybrid strategies.

Just as any other model, the approach in Kaznatcheev~et~al.\cite{Kaznatcheev2017a} is based on simplifying assumptions (e.g., only the subpopulation fractions are important, and no novel types can arise), which limit its practical applicability.
But our message is broader, and we use this specific model primarily to illustrate the general optimization approach.
With adaptive therapy trials now under way
showing early promise~\cite{Zhang2017a}, we propose that such methods will become increasingly integrated into trial design discussions, with increasingly detailed cancer evolution equations or even in data-driven/equation-free framework.

Of the two main approaches of optimal control theory, the Pontryagin Maximum Principle (PMP) has been much more widely used in cancer treatment research up till now.  
In contrast, our approach here is based on dynamic programming and the numerical methods for Hamilton-Jacobi-Bellman (HJB) equations.
The higher computational cost of these methods is balanced by several important practical considerations.
First, they yield a policy in feedback-form and are thus more robust to modeling/measurement errors.
Second, they always return the globally optimal treatment strategies and avoid some of the pitfalls well-know for the PMP-based methods (e.g., see Figure \ref{fig:ex0-b}).
With the advent of efficient numerical methods, we posit that the HJB equations will be soon playing a larger role in treatment optimization.

There are several obvious directions for extending our approach.
First, the ability to optimize outcomes
for a {\em range} of criteria will open new avenues to quantify physician/patient discussions concerning trade-offs and personalized therapy that have previously been only qualitative.
In the current paper, computing optimal policies for different values of $\sigma$ can be viewed as a small step in this direction.
But there are many other possible optimization criteria of practical interest.
Methods for approximating {\em all} Pareto-optimal policies
are available\cite{Kumar2010} but are usually more computationally challenging.
For probabilistic cancer evolution models, one can also choose between optimizing different characteristics of the same random quantity.  (E.g., minimize the average time-to-recovery versus maximizing the probability of recovery in the next year).
Finally, one can also use the choice of criterion to promote {\em robustness} by systematically treating possible measurement/modeling errors as perturbations chosen by some adversarial player.
Such ``games-against-nature'', as described recently by Stankova and colleagues~\cite{Stankova2018a}, can be similarly treated by solving Hamilton-Jacobi-Isaacs equations.

Another important limitation of the above techniques is our assumed full knowledge of the system state.
E.g., we assumed that the exact subpopulation fractions are known at every point in time and can be used to decide whether to administer the drugs.
In practice, one can periodically obtain an approximation of these quantities (e.g., based on a repeat biopsy), but most of the time the decisions must be made based on some less invasive measurements (e.g., based on PSA-levels in castrate-resistant prostate cancer model\cite{Zhang2017a} or in cell free circulating DNA as recently shown by Khan and colleagues~\cite{Khan2018}).
A rigorous treatment of such optimization challenges will require the framework of {\em partially-observable controlled processes}; see e.g., Davis and Varaiya\cite{Davis1973}.

With cancer evolution playing a larger and larger role in our thinking about therapy, and adaptive/evolutionary therapy coming to the fore, methods for optimizing these approaches will grow in importance.
We suggest that the HJB method be used in these scenarios, and that clinicians and modelers begin discussions of further method development along these lines.

\section*{Acknowledgement}
MG acknowledges the support from The Institute for Data and Decision Analytics  at The Chinese University of Hong Kong, Shenzhen during his visit when this research was completed.

JGS would like to thank the NIH Case Comprehensive Cancer Center support grant P30CA043703 and the Calabresi Clinical Oncology Research Program, National Cancer Institute of Award Number K12CA076917.

AV would like to thank the Simons Foundation for its fellowship support and the National Science Foundation (award DMS-1738010) for supporting development of numerical methods for Hamilton Jacobi equations.   A part of this work was performed during a sabbatical leave at Princeton/ORFE, and AV is grateful to ORFE Department for its hospitality.

\section{Conflict of Interest}

The authors declare no conflict of interest.

\section{Supplementary Materials}\label{SM}

\subsection{Deriving a Hamilton-Jacobi-Bellman equation}\label{HJB_sec}

We start by explaining the logic of dynamic programming that yields the HJB PDE~\eqref{IHJB} (Box~3, Section \ref{Quantifying_the_benefits}),
and the ``bang-bang'' property of optimal treatment policies.

Recall that the evolving composition of cancer sub-populations can be fully defined by $(q(t), p(t))$; see formulas \eqref{trans}
and \eqref{rec}. The process is tracked until we cross either a recovery or failure barrier; i.e., until the trajectory leaves $\Omega = \Big([0,1]\times[0,1]\Big) \backslash \Delta$, with the terminal set $\Delta$ defined in formula \eqref{ts}.  For an arbitrary initial state $(\qzero,\pzero)\in \Omega,$
the goal is to choose our treatment policy to minimize the integral of an instantaneous cost $K(d(t)) = d(t)+\sigma$ up to the terminal time $T=T(\qzero,\pzero, d(\cdot)).$
I.e., the total cost of starting at $(\qzero,\pzero)$ and using a policy $d(\cdot)$  is
\begin{equation*}
  J(\qzero,\pzero ,  d(\cdot) )\, = \, \int\limits_0^{T}
  K \left( d(s) \right) \, ds \, + \, g\Big(q\big( T\big), p\big( T\big)\Big),
\end{equation*}
where $g$ is the terminal cost specified on $\Delta$ in formula \eqref{tc}.
The {\em value function} $u(\qzero,\pzero)$ is the result of minimizing 
$J$
over all available treatment policies, and we say that the policy $d^*(\cdot)$ is optimal if $u(\qzero,\pzero) = J(\qzero,\pzero ,  d^*(\cdot))$.

Bellman's Optimality Principle \cite{Bellman1957} is the key idea of dynamic programming.  It states that, if we move along any optimal trajectory, a remaining (yet to be traversed) part of that trajectory is in itself optimal from our {\em current} configuration/state.  In terms of the above model,
\begin{equation}
\label{eq:Bellman}
u(\qzero,\pzero) \, = \, \int \limits_{0}^{\tau} K \left(d^*(t)\right) \, ds
\, + \, u\Big(q(\tau), p(\tau)\Big)
\end{equation}
should hold for every sufficiently small $\tau > 0.$
Assuming that the value function $u(q,p)$ and $d^*(t)$ are smooth, one can use Taylor series and take the limit $\tau\rightarrow 0$ to obtain
\begin{equation}\label{HJB_TS}
\nabla u(\qzero, \pzero)\cdot \left(
                                                                      \begin{array}{c}
                                                                        \dot q(\qzero,\pzero,d^*_0) \\
                                                                        \dot p(\qzero,\pzero,d^*_0) \\
                                                                      \end{array}
                                                                    \right) + d^*_0+\sigma = 0.
\end{equation}
Here $d^*_0 = d^*(0)$ is the optimal {\em initial} rate of therapy starting from $(\qzero,\pzero)$ and $(\dot q, \dot p)$ are specified by the right hand side of the ODEs in \eqref{rec}.  Since \eqref{HJB_TS} does not involve $d^*(t)$ for any $t>0$, it is now natural to switch to a {\em feedback control} perspective based on a state-dependent (rather than explicitly time-dependent) optimal control $d^*_0(q,p)$.  Since the latter is a priori unknown, a Hamilton-Jacobi PDE \eqref{IHJB} is obtained by minimizing over all available control values $d \in [0, \dmax]$ and demanding that \eqref{HJB_TS} should hold at every $(q,p)\in \Omega.$   Additional boundary conditions $u=g$ are specified on $\Delta$ by \eqref{bc}.

The above derivation is merely formal since the value function $u$ is typically non-smooth.  Indeed, \eqref{IHJB} rarely has classical solutions, and if one considers Lipschitz-continuous weak solutions (by demanding that the PDE should hold wherever $\nabla u$ is defined), one immediately loses the uniqueness.
Additional test conditions introduced by Crandall and Lions\cite{Crandall1983} are employed to pick out a {\em viscosity solution}  -- the unique weak solution coinciding with the value function of the original control problem \cite{Bardi1997}.   Convergence to this viscosity solution is also a requirement for all numerical methods for HJB equations used in control-theoretic applications.

Using the dynamics \eqref{rec} specific to our model, the HJB equation \eqref{IHJB} can be re-written as follows:
\begin{equation}
\min\limits_{d\in[0,\dmax]}
\left\{
~\Big[1- u^{}_p p(1-p)\Big] d \; + \; u^{}_q q(1-q)\Big(\frac{\bv}{n+1}\sum\limits_{k=0}^n p^k -c\Big)+u^{}_p p(1-p) \Big(\frac{\ba}{n+1} - q(\bv-c)\Big)  +\sigma~
\right\} \; = \; 0.
\end{equation}
The linear $d$-dependence of the minimized expression allows us to find the minimizer in closed form:
\begin{equation}\label{bang}
 d^*=
\begin{cases}
\dmax, & \quad \text{if } \Big(1- u^{}_p p(1-p)\Big) < 0;\\
0,  & \quad \text{otherwise}.
\end{cases}
\end{equation}
Therefore, an optimal treatment policy takes only extreme values -- either $0$ or $\dmax$. This is usually called the \emph{bang-bang} property.

Using \eqref{bang} in practice would require knowing $u^{}_p$ at every point $(q,p)$.
In principle $u^{}_p$, can be computed along an optimal trajectory (backwards, from the recovery barrier to our initial state $(\qzero,\pzero)$ without solving the PDE on the entire $\Omega$.  This is in a sense the main idea of Pontryagin Maximum Principle (PMP) \cite{Pontryagin1962}.  This method will work (and will be much more computationally attractive) as long as $u$ remains smooth along the optimal trajectory.  Unfortunately, the PMP has no way of identifying whether a backward-traced trajectory passes through a shockline (where $\nabla u$ is undefined).  In practice, this would result in obtaining locally (rather than globally) optimal treatment policies; see the example in Figure \ref{fig:ex0}c.
We thus focus on solving the full HJB equation, yielding a variational formula for $d^*(q,p).$

\subsection{Numerical methods for the Hamilton-Jacobi-Bellman equation}\label{TD}

\begin{figure}[H]
\centering
\subfigure[ ]{%
\label{fig:SL}
\includegraphics[height=2.5in]{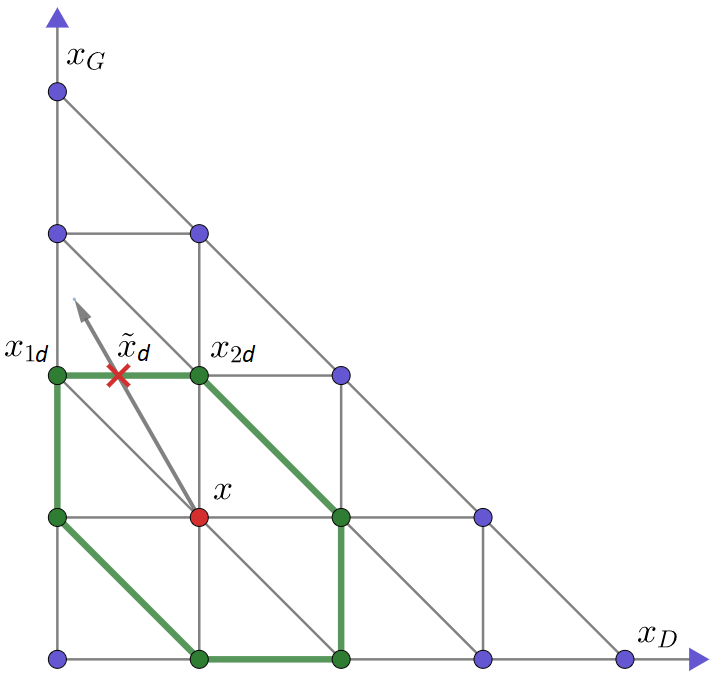}}%
\qquad
\subfigure[]{%
\label{fig:SL2}%
\includegraphics[height=2.5in]{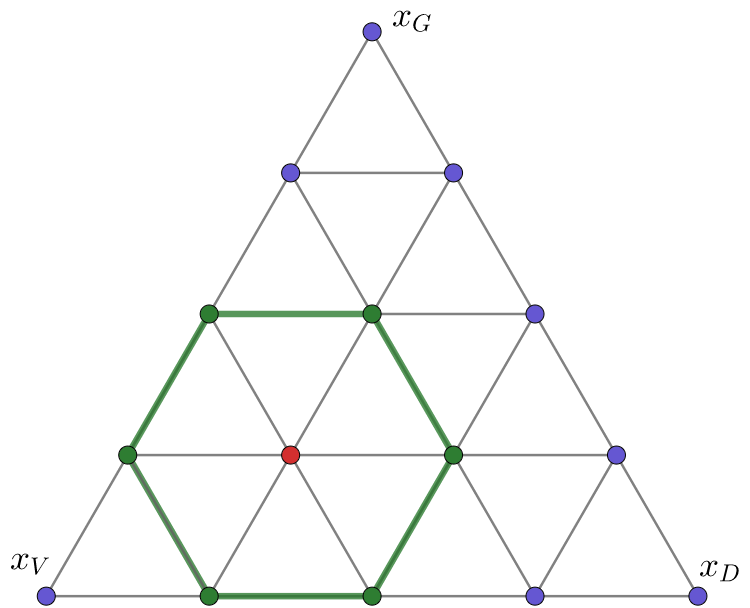}}%
\caption[]{\textbf{A Semi-Lagrangian scheme on a triangular mesh.}

\textbf{(a):} A semi-Lagrangian discretization in $(\xD, \xG).$

\textbf{(b):} Linear transformation yields a regular triangular mesh.
}\label{fig21}
\end{figure}

We obtain an approximate solution to HJB equations on a regular triangulated mesh over the $(\xD, \xG, \xV)$ space; see Figure \ref{fig:SL2}.
Since at any moment of time $ \xV(t) \equiv 1 - \xG(t) - \xD(t)$, it is enough to consider an ODE system for two sub-populations $\xD(t)$ and $\xG(t)$.  To simplify the notation, we will write

\begin{equation}\label{gen_ds}
\begin{cases}
\dot  y(t)= f\Big(y(t), d(t)\Big)\\
y(0) = x,
\end{cases}
\end{equation}
where $y(t) = (\xD(t), \xG(t))$, $x = (\xD(0), \xG(0))$ and $f(\cdot)$ denotes the right-hand side of \eqref{rec} in $(\xD, \xG)$ coordinates
using the transformation \eqref{trans}.

Our approximation scheme is based on a first-order accurate semi-Lagrangian discretization \cite{Falcone2013}.  Starting at a meshpoint $x$ and using control $d$, we assume that the rate of change is constant for a small time $\tau$, yielding a new state
 \begin{equation}\tilde{x}^{}_d = x+\tau f(x,d). \end{equation}
Assuming that the running cost is also constant over that small time interval,
one can rewrite Bellman's optimality principle
as
\begin{equation}
 u(x) \; = \; \min_d \left\{
 \tau K(d) \, + \, u \left( \tilde{x}^{}_d \right) \right\} \, + \, o(\tau).
\end{equation}
Since $\tilde{x}^{}_d$ is usually not a meshpoint, $u \left( \tilde{x}^{}_d \right)$ is approximated by interpolation using the neighboring meshpoint values.  (This is the key idea of all {\em semi-Lagrangian} techniques.)   While there are many ways to choose $\tau,$ we select it for each $d$ value individually
to guarantee that $\tilde{x}^{}_d$ lies on a mesh line and only two neighboring values are needed for the interpolation; a similar approach has been used in \cite{Gonzalez1985a, sethian2003ordered}.
More specifically, suppose that a vector $f(x,d)$ anchored at $x$ lies within a triangle $x \xoned\xtwod$; see Figure \ref{fig:SL}.  Then $\tilde{x}^{}_d$ lies on a segment $\xoned \xtwod$ with
$$
\tilde{x}^{}_d \; = \; \frac{||\xtwod-\tilde x^{}_d||}{||\xoned-\xtwod||} \xoned \, + \, \frac{||\xoned-\tilde x^{}_d||}{||\xoned-\xtwod||} \xtwod
\quad \text{ and } \quad
\tau^{}_d \, = \,  \frac{||x - \tilde x^{}_d ||}{||f(x,d)||}.
$$
Recalling that only extreme rates $d$ can be optimal due to the bang-bang property, we obtain a coupled system of discretized equations:
\begin{equation}\label{dvf}
U(x) \; = \; \min\limits_{d \in \{0,\, \dmax\}}\left\{
\frac{||x - \tilde x^{}_d ||}{||f(x,d)||}K(d) \, + \,
\frac{||\xtwod-\tilde x^{}_d||}{||\xoned-\xtwod||} U(\xoned) \, + \,
\frac{||\xoned-\tilde x^{}_d||}{||\xoned-\xtwod||} U(\xtwod)  \right\},
\end{equation}
which must hold for each meshpoint $x \in \Omega$.
The boundary conditions are handled by setting $U=0$ when $\xG < \rb$ and $U=+\infty$ when $\xG > 1- \fb$.

In our implementation, the above coupled system of discretized equations is handled by Gauss-Seidel iterations, with an additional speed up through  alternating meshpoint orderings (in a ``Fast Sweeping'' fashion)\cite{Boue1999, Zhao2004, Qian2007}.   Another alternative would be to decouple the system dynamically -- by selecting larger $\tau^{}_d$ adaptively so that ``already known'' mesh values would be sufficient
for updating the still-tentatively-known $U$ values.  The latter ``Ordered Upwind'' approach has been primarily used in problems with geometric dynamics \cite{sethian2003ordered, alton2012ordered, mirebeau2014efficient} and offers advantages when optimal trajectories frequently change directions. In the future, it would be interesting to extend it (as well as its two-scale hybrids with sweeping  \cite{Chacon2012}) to therapy optimization problems, particularly for the case of small $\sigma$.

From implementation purposes, it is easier to conduct computations in a Cartesian coordinate system (Figure \ref{fig:SL}), which is equivalent to a regular triangular mesh (Figure \ref{fig:SL2}) by a linear transformation.

To ensure the accuracy of the value function in Figure \ref{fig:ex0-a}, we have used $n=9000$ meshpoints along one side of the GLY-VOP-DEF triangle, yielding $N=37~988~686$ meshpoints in $\Omega$ with $\rb=\fb = 10^{-1.5}.$
The algorithm terminates when the difference between value functions in sequential iterations
falls below $10^{-5}$, which required a total of 62 iterations (sweeps) for this example.

In the future, we hope to reduce the computational cost by using a higher-order accurate semi-Lagrangian discretization \cite{Falcone2013} on a coarser mesh, employing ``Ordered Upwind''  techniques to reduce or eliminate the coupling in the discretized system.

\subsection{Fully angiogenic and glycolyctic tumors}\label{other_regimes}

In section \ref{results} we have focused on optimizing treatment policies for polyclonal tumors. Under ``therapy-off'' policy any trajectory of a polyclonal tumour has  periodic dynamics and model parameters satisfy (\ref{cyc}).
Two other types of tumours are also possible under the model\cite{Kaznatcheev2017a}: fully angiogenic and fully glycolyctic.

\begin{figure}[H]
\centering

\subfigure[][]{%
\label{fig:ex10-a}%
\includegraphics[height=2in]{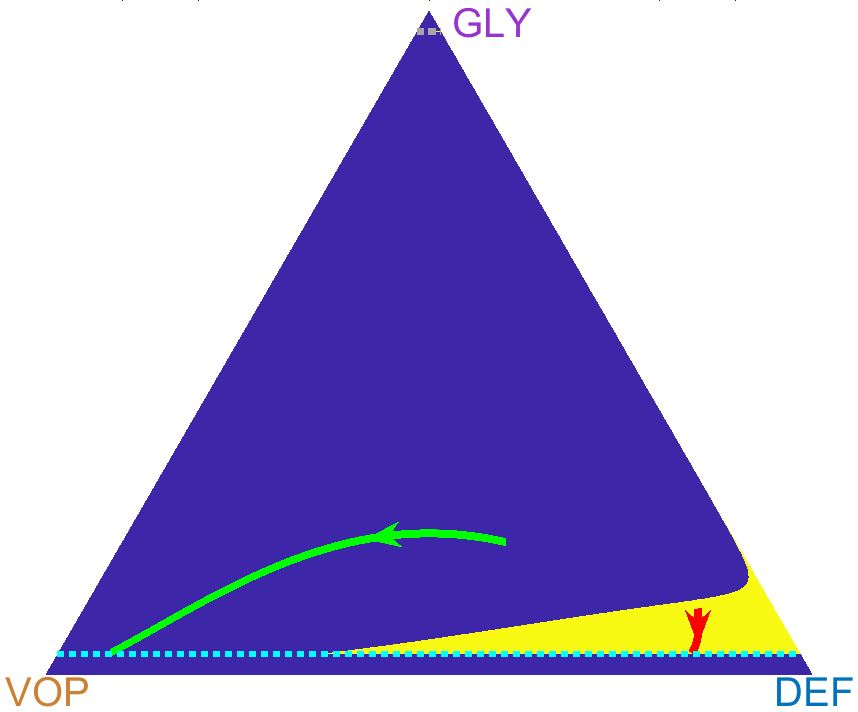}}%
\hspace{20pt}%
\subfigure[][]{%
\label{fig:ex10-b}%
\includegraphics[height=2in]{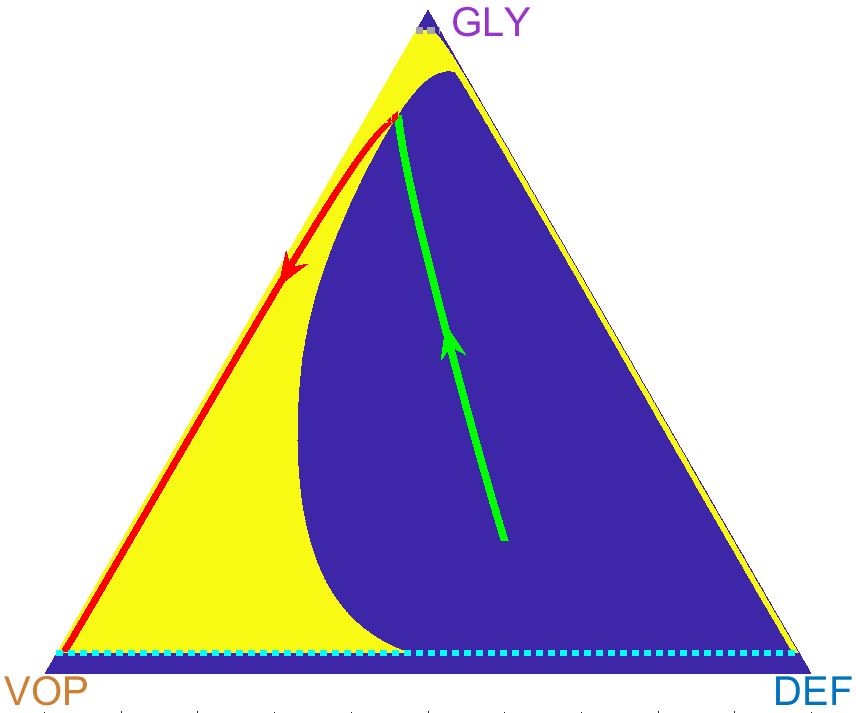}}

\caption[]{\textbf{
The benefits of optimization in fully angiogenic and glycolyctic cases.}

 \textbf{\subref{fig:ex10-a}}: two trajectories under AT strategy in a \textit{fully angiogenic} tumour.
 Parameters:  $\ba = 2$, $\bv=3$, $c=1$, $n=1$; $\dmax=3$, $\sigma = 0.3$, $\rb=\fb = 10^{-1.5}$.

 \textbf{\subref{fig:ex10-b}}: a trajectory under AT strategies in a \textit{fully glycolyctic} tumour.
 Parameters: $\ba = 30$, $\bv=6$, $c=1$, $n=4$; $\dmax=3$, $\sigma = 0.01$, $\rb=\fb = 10^{-1.5}$.
 }
\label{fig:ex10}%
\end{figure}

A tumor has a fully angiogenic regime if $\max\Big(\frac{\ba}{n+1}, cn \Big)<\bv-c$.
If  the model (\ref{re2d}) satisfies this condition all cells tend to switch to VOP type and the trajectory converges to the recovery zone\cite{Kaznatcheev2017a}.
In some sense, the fully angiogenic regime is less interesting case for our analysis because even without any therapy a patient will recover.
However, the optimal control analysis still might be useful when, for example, time-penalty is high (a patient wants to recover as soon as possible) and some amount of drugs can be applied to accelerate the recovery, see Figure \ref{fig:ex10-a}.

A tumor has a fully glycolyctic regime if $\frac{\ba}{n+1}>\bv-c$, all cells tend to be GLY type cells and  trajectories converge to the failure zone from any initial state.
Even if a treatment policy gives some short-term results,
the trajectory will turn towards the failure zone once the therapy is  stopped.
Nevertheless, crossing the recovery barrier means  full recovery under assumptions of the model\cite{Kaznatcheev2017a}. We consider an example of optimal policy for fully glycolyctic tumour in Figure \ref{fig:ex10-b}.

\subsection{Incurable areas changing under parameter variation}
\label{ss:incurable_param_var}

\begin{figure}[H]
\begin{multicols}{2}
    \centering%

\settoheight{\tempdima}{\includegraphics[width=.25\linewidth]{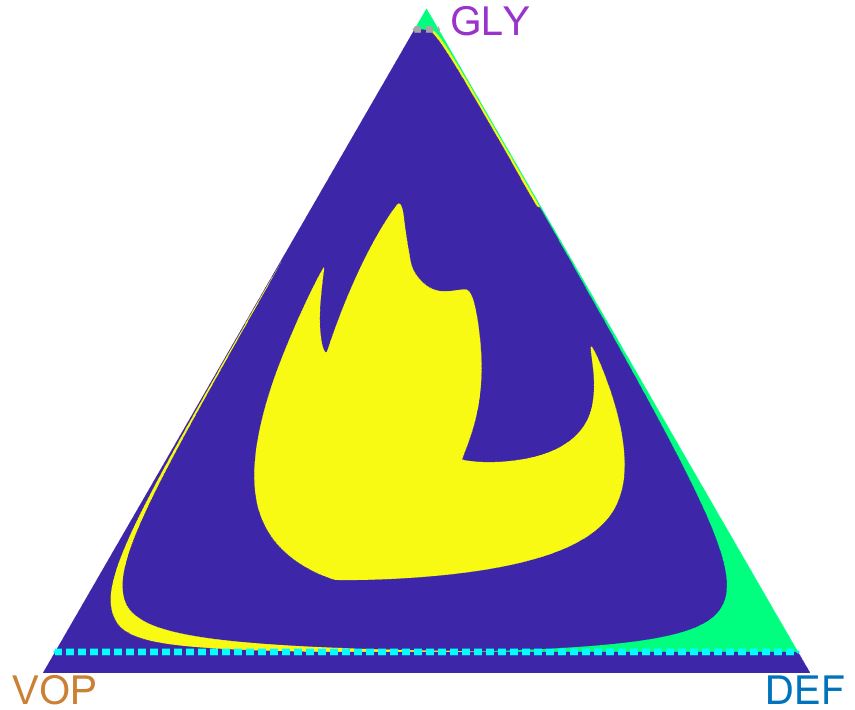}}%
\centering\begin{tabular}{@{}c@{ }c@{ }c@{ }c@{}}
\subfigure[]{%
\label{fig:ex8-a}%
\includegraphics[height=1.2in]{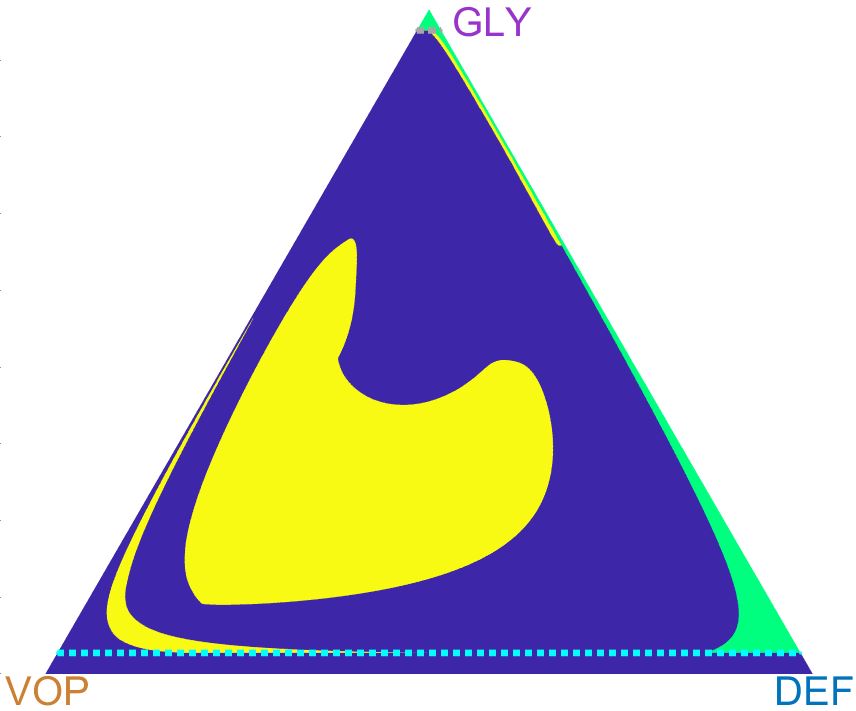}}%
\hspace{8pt}%
\subfigure[]{%
\label{fig:ex8-b}%
\includegraphics[height=1.2in]{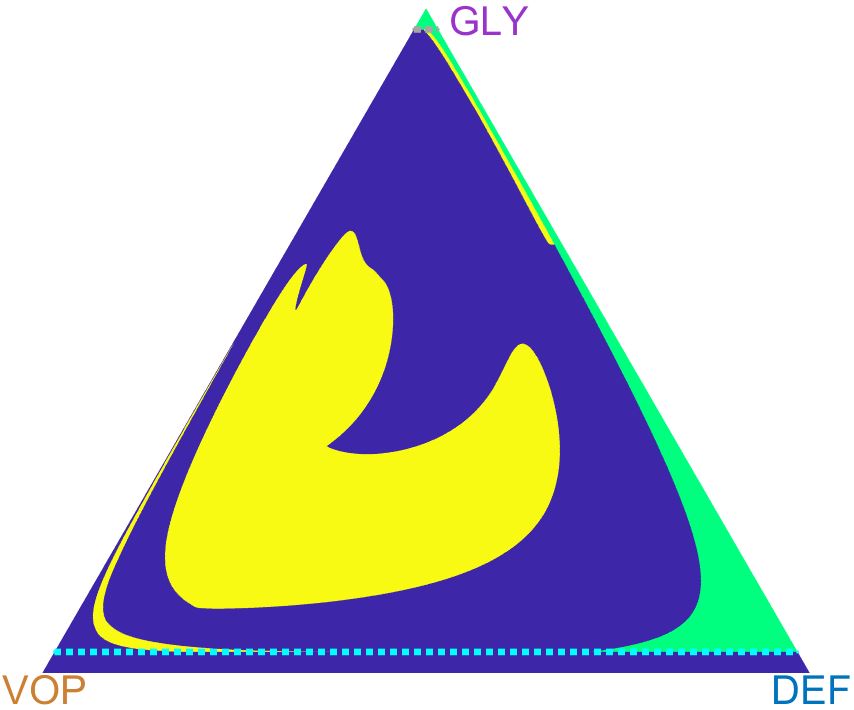}}%
\hspace{8pt}%
\subfigure[]{%
\label{fig:ex8-c}%
\includegraphics[height=1.2in]{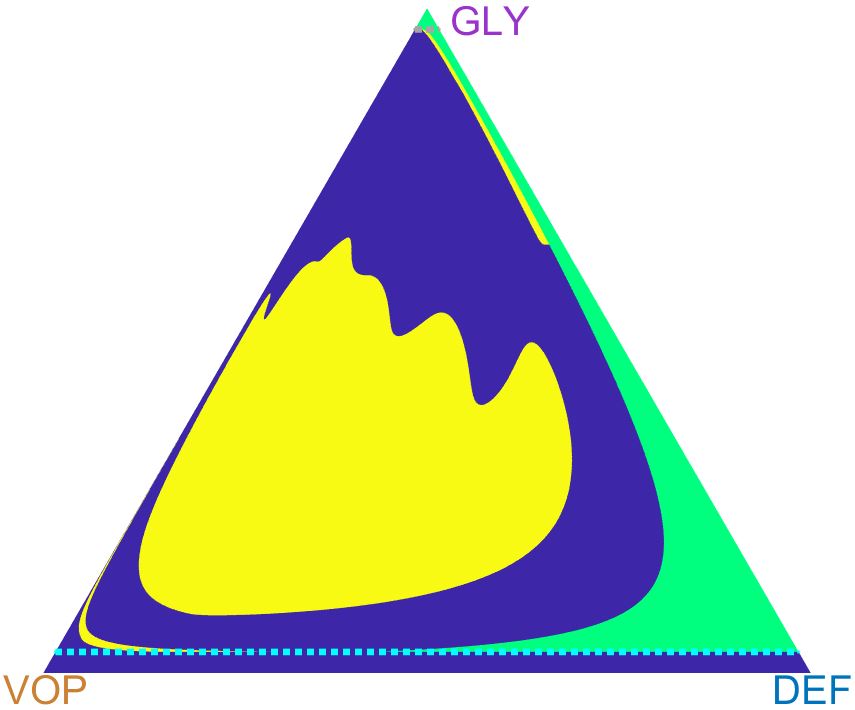}}\\
\hspace{8pt}%
\subfigure[]{%
\label{fig:ex8-d}%
\includegraphics[height=1.2in]{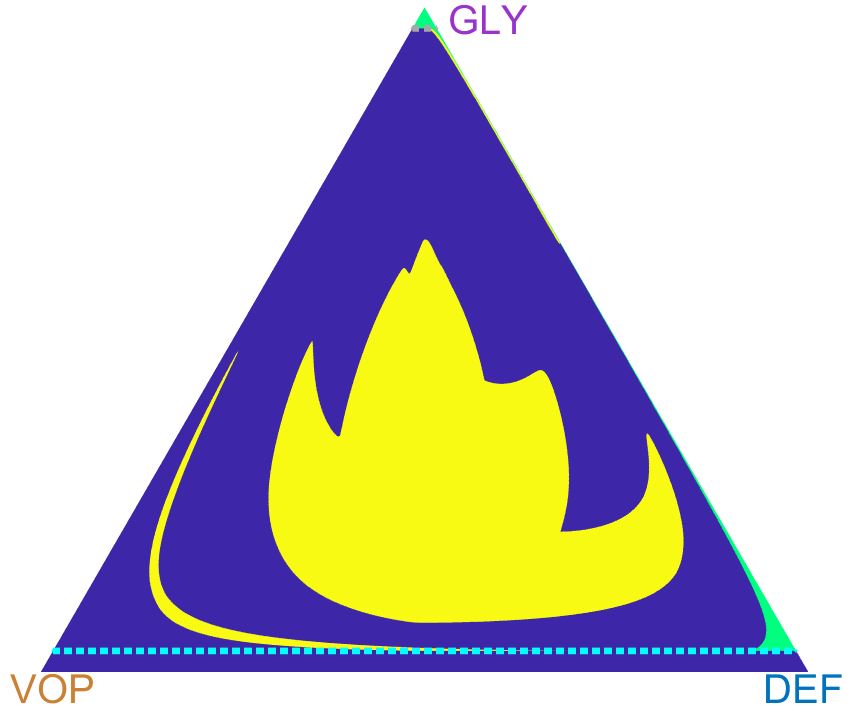}}
\subfigure[]{%
\label{fig:ex8-e}%
\includegraphics[height=1.2in]{plot_ha_bv2.JPG}}
\hspace{8pt}%
\subfigure[]{%
\label{fig:ex8-f}%
\includegraphics[height=1.2in]{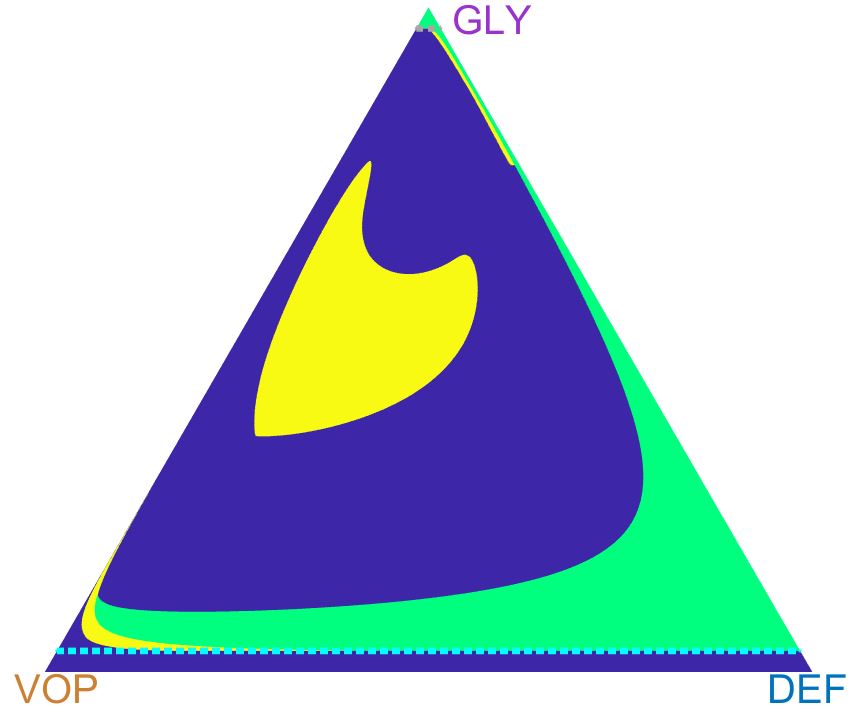}}

\end{tabular}\par~\vspace{0.8mm}\par

    {\small
\centering
\renewcommand{\arraystretch}{1.8}
\hspace{28mm}\begin{tabular}{
!{\vrule width 2pt}c!{\vrule width 2pt}c|c|c|c!{\vrule width 2pt}}
\ChangeRT{2pt}
\multirow{2}{*}{Subfigure}&\multicolumn{4}{c!{\vrule width 2pt}}{Parameters} \\
\cline{2-5}
&$\dmax$ & $\sigma$ & $\ba$ & $\bv$   \\
\ChangeRT{2pt}
\subref{fig:ex8-a}& $\textbf{0.3}$ &$0.03$ & $4$&$2$  \\
\hline
\subref{fig:ex8-b}& $\textbf{0.2}$ &$0.03$ &$4$&$2$  \\
\hline
\subref{fig:ex8-c}& $\textbf{0.1}$ &$0.03$ &$4$&$2$ \\
\ChangeRT{2pt}
\subref{fig:ex8-d}& $0.2$ &$0.01$ & $2.5$&$\textbf{2}$ \\
\hline
\subref{fig:ex8-e}& $0.2$ &$0.01$ & $2.5$&$\textbf{1.8}$  \\
\hline

\subref{fig:ex8-f}& $ 0.2$ &$0.01$ &$2.5$&$\textbf{1.6}$  \\
\ChangeRT{2pt}
\end{tabular}
}
    \end{multicols}

    \caption[]{ \textbf{Optimal drugs-on regions (in yellow) and incurable areas (in green) changing under parameter variation.}
    
    ``Incurable'' areas can grow due to a decrease in the MTD rate $\dmax$ (top row) or a decrease in vascularization benefits $\bv$ (bottom row). Common parameters for the figures: $c=1$, $n=4$, $\rb=\fb=10^{-1.5}$.  }%
\label{fig:ex8}%
\end{figure}

In Figure \ref{fig:ex8} we examine the changes in the optimal/minimal incurable area due to variations in the MTD rate and model parameters.

\bibliographystyle{unsrtnat}
\bibliography{Cancer_treatment_optimal_scheduling}

\begin{thebibliography}{57}
\providecommand{\natexlab}[1]{#1}
\providecommand{\url}[1]{\texttt{#1}}
\expandafter\ifx\csname urlstyle\endcsname\relax
  \providecommand{\doi}[1]{doi: #1}\else
  \providecommand{\doi}{doi: \begingroup \urlstyle{rm}\Url}\fi

\bibitem[Marusyk and Polyak(2010)]{Marusyk2010}
Andriy Marusyk and Kornelia Polyak.
\newblock {Tumor heterogeneity: Causes and consequences}.
\newblock \emph{Biochimica et Biophysica Acta (BBA) - Reviews on Cancer},
  1805\penalty0 (1):\penalty0 105--117, 2010.

\bibitem[Hanahan and Weinberg(2000)]{Hanahan2000}
D~Hanahan and R~A Weinberg.
\newblock {The hallmarks of cancer.}
\newblock \emph{Cell}, 100\penalty0 (1):\penalty0 57--70, 2000.

\bibitem[Scott and Marusyk(2017)]{Scott2017}
Jacob Scott and Andriy Marusyk.
\newblock {Somatic clonal evolution: A selection-centric perspective}.
\newblock \emph{Biochimica et Biophysica Acta (BBA) - Reviews on Cancer},
  1867\penalty0 (2):\penalty0 139--150, 2017.
\newblock \doi{10.1016/J.BBCAN.2017.01.006}.

\bibitem[Marusyk et~al.(2014)Marusyk, Tabassum, Altrock, Almendro, Michor, and
  Polyak]{Marusyk2014}
Andriy Marusyk, Doris~P. Tabassum, Philipp~M. Altrock, Vanessa Almendro,
  Franziska Michor, and Kornelia Polyak.
\newblock {Non-cell-autonomous driving of tumour growth supports sub-clonal
  heterogeneity}.
\newblock \emph{Nature}, 514\penalty0 (7520):\penalty0 54--58, 2014.
\newblock \doi{10.1038/nature13556}.

\bibitem[Hanahan et~al.(2000)Hanahan, Bergers, and Bergsland]{Hanahan2000a}
Douglas Hanahan, Gabriele Bergers, and Emily Bergsland.
\newblock {Less is more, regularly: metronomic dosing of cytotoxic drugs can
  target tumor angiogenesis in mice}.
\newblock \emph{Journal of Clinical Investigation}, 105\penalty0 (8):\penalty0
  1045--1047, 2000.

\bibitem[Lien et~al.(2013)Lien, Georgsdottir, Sivanathan, Chan, and
  Emmenegger]{Lien2013}
K.~Lien, S.~Georgsdottir, L.~Sivanathan, K.~Chan, and U.~Emmenegger.
\newblock {Low-dose metronomic chemotherapy: A systematic literature analysis}.
\newblock \emph{European Journal of Cancer}, 49\penalty0 (16):\penalty0
  3387--3395, 2013.
\newblock \doi{10.1016/J.EJCA.2013.06.038}.

\bibitem[Pasquier et~al.(2010)Pasquier, Kavallaris, and
  Andr{\'{e}}]{Pasquier2010}
Eddy Pasquier, Maria Kavallaris, and Nicolas Andr{\'{e}}.
\newblock {Metronomic chemotherapy: new rationale for new directions}.
\newblock \emph{Nature Reviews Clinical Oncology}, 7\penalty0 (8):\penalty0
  455--465, 2010.
\newblock \doi{10.1038/nrclinonc.2010.82}.

\bibitem[Kesari et~al.(2007)Kesari, Schiff, Doherty, Gigas, Batchelor, and
  Muzikansky~et al.]{Kesari2007}
Santosh Kesari, David Schiff, Lisa Doherty, Debra~C Gigas, Tracy~T Batchelor,
  and Alona Muzikansky~et al.
\newblock {Phase II study of metronomic chemotherapy for recurrent malignant
  gliomas in adults.}
\newblock \emph{Neuro-oncology}, 9\penalty0 (3):\penalty0 354--363, 2007.

\bibitem[Steinbild et~al.(2007)Steinbild, Arends, Medinger, H{\"{a}}ring,
  Frost, Drevs, Unger, Strecker, Hennig, and Mross]{Steinbild2007}
Simone Steinbild, Jann Arends, Michael Medinger, Brigitte H{\"{a}}ring, Annette
  Frost, Joachim Drevs, Clemens Unger, Ralph Strecker, J{\"{u}}rgen Hennig, and
  Klaus Mross.
\newblock {Metronomic Antiangiogenic Therapy with Capecitabine and Celecoxib in
  Advanced Tumor Patients - Results of a Phase II Study}.
\newblock \emph{Oncology Research and Treatment}, 30\penalty0 (12):\penalty0
  629--635, 2007.

\bibitem[Senerchia et~al.(2017)Senerchia, Macedo, Ferman, Scopinaro,
  Cacciavillano, and Boldrini~et al.]{Senerchia2017}
Andreza~A. Senerchia, Carla~Renata Macedo, Sima Ferman, Marcelo Scopinaro,
  Walter Cacciavillano, and Erica Boldrini~et al.
\newblock {Results of a randomized, prospective clinical trial evaluating
  metronomic chemotherapy in nonmetastatic patients with high-grade, operable
  osteosarcomas of the extremities: A report from the Latin American Group of
  Osteosarcoma Treatment}.
\newblock \emph{Cancer}, 123\penalty0 (6):\penalty0 1003--1010, 2017.

\bibitem[Chen et~al.(2014)Chen, Doloff, and Waxman]{Chen2014}
Chong-Sheng Chen, Joshua~C. Doloff, and David~J. Waxman.
\newblock {Intermittent Metronomic Drug Schedule Is Essential for Activating
  Antitumor Innate Immunity and Tumor Xenograft Regression}.
\newblock \emph{Neoplasia}, 16\penalty0 (1):\penalty0 84--96, 2014.
\newblock \doi{10.1593/NEO.131910}.

\bibitem[Gatenby et~al.(2009)Gatenby, Silva, Gillies, and Frieden]{Gatenby2009}
R.~A. Gatenby, A.~S. Silva, R.~J. Gillies, and B.~R. Frieden.
\newblock {Adaptive Therapy}.
\newblock \emph{Cancer Research}, 69\penalty0 (11):\penalty0 4894--4903, 2009.

\bibitem[Enriquez-Navas et~al.(2016)Enriquez-Navas, Kam, Das, Hassan, Silva,
  and Foroutan~et al.]{Enriquez-Navas2016}
Pedro~M Enriquez-Navas, Yoonseok Kam, Tuhin Das, Sabrina Hassan, Ariosto Silva,
  and Parastou Foroutan~et al.
\newblock {Exploiting evolutionary principles to prolong tumor control in
  preclinical models of breast cancer.}
\newblock \emph{Science translational medicine}, 8\penalty0 (327):\penalty0
  327ra24, 2016.

\bibitem[Zhang et~al.(2017)Zhang, Cunningham, Brown, and Gatenby]{Zhang2017a}
Jingsong Zhang, Jessica~J. Cunningham, Joel~S. Brown, and Robert~A. Gatenby.
\newblock {Integrating evolutionary dynamics into treatment of metastatic
  castrate-resistant prostate cancer}.
\newblock \emph{Nature Communications}, 8\penalty0 (1):\penalty0 1816, 2017.
\newblock \doi{10.1038/s41467-017-01968-5}.

\bibitem[Smith(1982)]{Smith1982}
John~Maynard Smith.
\newblock \emph{{Evolution and the theory of games}}.
\newblock Cambridge University Press, Cambridge, 1982.
\newblock ISBN 9780511806292.
\newblock \doi{10.1017/CBO9780511806292}.

\bibitem[Hofbauer and Sigmund(1998)]{Hofbauer1998a}
Josef Hofbauer and Karl Sigmund.
\newblock \emph{{Evolutionary games and population dynamics}}.
\newblock Cambridge University Press, 1998.
\newblock ISBN 9780521625708.

\bibitem[Basanta et~al.(2012)Basanta, Scott, Fishman, Ayala, Hayward, and
  Anderson]{Basanta2012}
D~Basanta, J~G Scott, M~N Fishman, G~Ayala, S~W Hayward, and A~R~A Anderson.
\newblock {Investigating prostate cancer tumour-stroma interactions: clinical
  and biological insights from an evolutionary game}.
\newblock \emph{British Journal of Cancer}, 106\penalty0 (1):\penalty0
  174--181, 2012.
\newblock \doi{10.1038/bjc.2011.517}.

\bibitem[You et~al.(2017)You, Brown, Thuijsman, Cunningham, Gatenby, and
  Zhang~et al.]{You2017}
Li~You, Joel~S. Brown, Frank Thuijsman, Jessica~J. Cunningham, Robert~A.
  Gatenby, and Jingsong Zhang~et al.
\newblock {Spatial vs. non-spatial eco-evolutionary dynamics in a tumor growth
  model}.
\newblock \emph{Journal of Theoretical Biology}, 435:\penalty0 78--97, 2017.
\newblock \doi{10.1016/j.jtbi.2017.08.022}.

\bibitem[Cunningham et~al.(2018)Cunningham, Brown, Gatenby, and
  Staňkov{\'{a}}]{Cunningham2018}
Jessica~J. Cunningham, Joel~S. Brown, Robert~A. Gatenby, and Kateřina
  Staňkov{\'{a}}.
\newblock {Optimal control to develop therapeutic strategies for metastatic
  castrate resistant prostate cancer}.
\newblock \emph{Journal of Theoretical Biology}, 459:\penalty0 67--78, 2018.

\bibitem[Basanta et~al.(2011)Basanta, Scott, Rockne, Swanson, and
  Anderson]{Basanta2011a}
David Basanta, Jacob~G Scott, Russ Rockne, Kristin~R Swanson, and Alexander R~A
  Anderson.
\newblock {The role of IDH1 mutated tumour cells in secondary glioblastomas: an
  evolutionary game theoretical view}.
\newblock \emph{Physical Biology}, 8\penalty0 (1):\penalty0 015016, 2011.
\newblock \doi{10.1088/1478-3975/8/1/015016}.

\bibitem[Dingli et~al.(2009)Dingli, Offord, Myers, Peng, Carr, and Josic~et
  al.]{Dingli2009}
D~Dingli, C~Offord, R~Myers, K-W Peng, T~W Carr, and K~Josic~et al.
\newblock {Dynamics of multiple myeloma tumor therapy with a recombinant
  measles virus.}
\newblock \emph{Cancer gene therapy}, 16\penalty0 (12):\penalty0 873--82, 2009.
\newblock \doi{10.1038/cgt.2009.40}.

\bibitem[Wu et~al.(2014)Wu, Liao, Tlsty, Sturm, and Austin]{Wu2014}
A.~Wu, D.~Liao, T.~D. Tlsty, J.~C. Sturm, and R.~H. Austin.
\newblock {Game theory in the death galaxy: interaction of cancer and stromal
  cells in tumour microenvironment}.
\newblock \emph{Interface Focus}, 4\penalty0 (4):\penalty0 20140028, 2014.
\newblock \doi{10.1098/rsfs.2014.0028}.

\bibitem[Komarova and Wodarz(2005)]{Komarova2005}
N.~L. Komarova and D.~Wodarz.
\newblock {Drug resistance in cancer: Principles of emergence and prevention}.
\newblock \emph{Proceedings of the National Academy of Sciences}, 102\penalty0
  (27):\penalty0 9714--9719, 2005.
\newblock \doi{10.1073/pnas.0501870102}.

\bibitem[Orlando et~al.(2012)Orlando, Gatenby, and Brown]{Orlando2012}
Paul~A Orlando, Robert~A Gatenby, and Joel~S Brown.
\newblock {Cancer treatment as a game: integrating evolutionary game theory
  into the optimal control of chemotherapy.}
\newblock \emph{Physical biology}, 9\penalty0 (6):\penalty0 065007, 2012.

\bibitem[West et~al.(2016)West, Hasnain, Mason, and Newton]{West2016}
Jeffrey West, Zaki Hasnain, Jeremy Mason, and Paul~K Newton.
\newblock {The prisoner's dilemma as a cancer model}.
\newblock \emph{Convergent Science Physical Oncology}, 2\penalty0 (3):\penalty0
  035002, 2016.
\newblock \doi{10.1088/2057-1739/2/3/035002}.

\bibitem[Sch{\"{a}}ttler and Ledzewicz(2015)]{Schattler2015}
Heinz Sch{\"{a}}ttler and Urszula Ledzewicz.
\newblock \emph{{Optimal Control for Mathematical Models of Cancer Therapies}},
  volume~42 of \emph{Interdisciplinary Applied Mathematics}.
\newblock Springer New York, New York, NY, 2015.
\newblock ISBN 978-1-4939-2971-9.
\newblock \doi{10.1007/978-1-4939-2972-6}.

\bibitem[Swan and Vincent(1977)]{Swan1977}
George~W. Swan and Thomas~L. Vincent.
\newblock {Optimal control analysis in the chemotherapy of IgG multiple
  myeloma}.
\newblock \emph{Bulletin of Mathematical Biology}, 39\penalty0 (3):\penalty0
  317--337, 1977.

\bibitem[Pontryagin et~al.(1962)Pontryagin, Boltyanskii, Gamkrelidze, and
  Mishchenko]{Pontryagin1962}
L.~Pontryagin, V.~Boltyanskii, R.~Gamkrelidze, and E.~Mishchenko.
\newblock \emph{{The mathematical theory of optimal processes}}.
\newblock John Wiley {\&} Sons, Inc., New York, 1962.

\bibitem[Schattler and Ledzewicz(2006)]{Schattler2006}
Heinz Schattler and Urszula Ledzewicz.
\newblock {Drug resistance in cancer chemotherapy as an optimal control
  problem}.
\newblock \emph{Discrete and Continuous Dynamical Systems - Series B},
  6\penalty0 (1):\penalty0 129--150, 2006.
\newblock \doi{10.3934/dcdsb.2006.6.129}.

\bibitem[Wang and Sch{\"{a}}ttler(2016)]{Wang2016}
Shuo Wang and Heinz Sch{\"{a}}ttler.
\newblock {Optimal control of a mathematical model for cancer chemotherapy
  under tumor heterogeneity}.
\newblock \emph{Mathematical Biosciences and Engineering}, 13\penalty0
  (6):\penalty0 1223--1240, 2016.
\newblock \doi{10.3934/mbe.2016040}.

\bibitem[Carr{\`{e}}re(2017)]{Carrere2017}
C{\'{e}}cile Carr{\`{e}}re.
\newblock {Optimization of an in vitro chemotherapy to avoid resistant
  tumours}.
\newblock \emph{Journal of Theoretical Biology}, 413:\penalty0 24--33, 2017.
\newblock \doi{10.1016/J.JTBI.2016.11.009}.

\bibitem[D'Onofrio et~al.(2009)D'Onofrio, Ledzewicz, Maurer, and
  Sch{\"{a}}ttler]{DOnofrio2009}
Alberto D'Onofrio, Urszula Ledzewicz, Helmut Maurer, and Heinz Sch{\"{a}}ttler.
\newblock {On optimal delivery of combination therapy for tumors}.
\newblock \emph{Mathematical Biosciences}, 222\penalty0 (1):\penalty0 13--26,
  2009.
\newblock \doi{10.1016/J.MBS.2009.08.004}.

\bibitem[Su et~al.(2016)Su, Jia, and Chen]{Su2016}
Yongmei Su, Chen Jia, and Ying Chen.
\newblock {Optimal Control Model of Tumor Treatment with Oncolytic Virus and
  MEK Inhibitor.}
\newblock \emph{BioMed research international}, 2016:\penalty0 5621313, 2016.

\bibitem[Ledzewicz et~al.(2012)Ledzewicz, Naghnaeian, and
  Sch{\"{a}}ttler]{Ledzewicz2012}
Urszula Ledzewicz, Mohammad Naghnaeian, and Heinz Sch{\"{a}}ttler.
\newblock {Optimal response to chemotherapy for a mathematical model of
  tumor-immune dynamics}.
\newblock \emph{Journal of Mathematical Biology}, 64\penalty0 (3):\penalty0
  557--577, 2012.

\bibitem[Lemos et~al.(2016)Lemos, Caiado, Coelho, and Vinga]{Lemos2016}
Jo{\~{a}}o~M. Lemos, Daniela~V. Caiado, Rui Coelho, and Susana Vinga.
\newblock {Optimal and receding horizon control of tumor growth in myeloma bone
  disease}.
\newblock \emph{Biomedical Signal Processing and Control}, 24:\penalty0
  128--134, 2016.
\newblock \doi{10.1016/J.BSPC.2015.10.004}.

\bibitem[Bardi and Dolcetta(1997)]{Bardi1997}
Martino Bardi and Italo Dolcetta.
\newblock \emph{{Optimal Control and Viscosity Solutions of
  Hamilton-Jacobi-Bellman Equations}}.
\newblock Birkh{\"{a}}user, Boston, MA, 1997.

\bibitem[Liberzon(2012)]{Liberzon2012}
Daniel Liberzon.
\newblock \emph{{Calculus of variations and optimal control theory : a concise
  introduction}}.
\newblock Princeton University Press, Princeton, Oxford, 2012.
\newblock ISBN 0691151873.

\bibitem[Nowakowski and Popa(2013)]{Nowakowski2013}
A.~Nowakowski and A.~Popa.
\newblock {A Dynamic Programming Approach for Approximate Optimal Control for
  Cancer Therapy}.
\newblock \emph{Journal of Optimization Theory and Applications}, 156\penalty0
  (2):\penalty0 365--379, 2013.
\newblock \doi{10.1007/s10957-012-0137-z}.

\bibitem[Lorz et~al.(2013)Lorz, Lorenzi, Hochberg, Clairambault, and
  Perthame]{Lorz2013}
Alexander Lorz, Tommaso Lorenzi, Michael~E. Hochberg, Jean Clairambault, and
  Beno{\^{i}}t Perthame.
\newblock {Populational adaptive evolution, chemotherapeutic resistance and
  multiple anti-cancer therapies}.
\newblock \emph{ESAIM: Mathematical Modelling and Numerical Analysis},
  47\penalty0 (2):\penalty0 377--399, 2013.

\bibitem[Kaznatcheev et~al.(2017{\natexlab{a}})Kaznatcheev, {Vander Velde},
  Scott, and Basanta]{Kaznatcheev2017a}
Artem Kaznatcheev, Robert {Vander Velde}, Jacob~G Scott, and David Basanta.
\newblock {Cancer treatment scheduling and dynamic heterogeneity in social
  dilemmas of tumour acidity and vasculature}.
\newblock \emph{British Journal of Cancer}, 116\penalty0 (6):\penalty0
  785--792, 2017{\natexlab{a}}.

\bibitem[Kaznatcheev et~al.(2017{\natexlab{b}})Kaznatcheev, Peacock, Basanta,
  Marusyk, and Scott]{Kaznatcheev2017b}
Artem Kaznatcheev, Jeffrey Peacock, David Basanta, Andriy Marusyk, and Jacob~G.
  Scott.
\newblock {Fibroblasts and alectinib switch the evolutionary games that
  non-small cell lung cancer plays}.
\newblock \emph{bioRxiv}, 2017{\natexlab{b}}.
\newblock \doi{10.1101/179259}.
\newblock URL \url{https://www.biorxiv.org/content/early/2017/09/20/179259}.

\bibitem[Stankov{\'{a}} et~al.(2018)Stankov{\'{a}}, Brown, Dalton, and
  Gatenby]{Stankova2018a}
Katerina Stankov{\'{a}}, Joel~S. Brown, William~S. Dalton, and Robert~A.
  Gatenby.
\newblock {Optimizing Cancer Treatment Using Game Theory}.
\newblock \emph{JAMA Oncology}, 2018.
\newblock \doi{10.1001/jamaoncol.2018.3395}.
\newblock URL
  \url{http://oncology.jamanetwork.com/article.aspx?doi=10.1001/jamaoncol.2018.3395}.

\bibitem[Nichol et~al.(2015)Nichol, Jeavons, Fletcher, Bonomo, Maini, and
  Paul~et al.]{Nichol2015}
Daniel Nichol, Peter Jeavons, Alexander~G. Fletcher, Robert~A. Bonomo,
  Philip~K. Maini, and Jerome~L. Paul~et al.
\newblock {Steering Evolution with Sequential Therapy to Prevent the Emergence
  of Bacterial Antibiotic Resistance}.
\newblock \emph{PLOS Computational Biology}, 11\penalty0 (9):\penalty0
  e1004493, 2015.
\newblock \doi{10.1371/journal.pcbi.1004493}.

\bibitem[Kumar and Vladimirsky(2010)]{Kumar2010}
Ajeet Kumar and Alexander Vladimirsky.
\newblock {An efficient method for multiobjective optimal control and optimal
  control subject to integral constraints}.
\newblock \emph{Journal of Computational Mathematics}, 28:\penalty0 517--551,
  2010.
\newblock \doi{10.2307/43693920}.

\bibitem[Khan et~al.(2018)Khan, Cunningham, Werner, Vlachogiannis, Spiteri, and
  Heide~et al.]{Khan2018}
Khurum~H Khan, David Cunningham, Benjamin Werner, Georgios Vlachogiannis,
  Inmaculada Spiteri, and Timon Heide~et al.
\newblock {Longitudinal Liquid Biopsy and Mathematical Modeling of Clonal
  Evolution Forecast Time to Treatment Failure in the PROSPECT-C Phase II
  Colorectal Cancer Clinical Trial.}
\newblock \emph{Cancer discovery}, 8\penalty0 (10):\penalty0 1270--1285, 2018.
\newblock \doi{10.1158/2159-8290.CD-17-0891}.

\bibitem[Davis and Varaiya(1973)]{Davis1973}
M.~H.~A. Davis and P.~Varaiya.
\newblock {Dynamic Programming Conditions for Partially Observable Stochastic
  Systems}.
\newblock \emph{SIAM Journal on Control}, 11\penalty0 (2):\penalty0 226--261,
  1973.

\bibitem[Bellman(1957)]{Bellman1957}
Richard Bellman.
\newblock \emph{{Dynamic programming}}.
\newblock Princeton University Press, 1957.
\newblock ISBN 0486428095.

\bibitem[Crandall and Lions(1983)]{Crandall1983}
Michael~G. Crandall and Pierre-Louis Lions.
\newblock {Viscosity solutions of Hamilton-Jacobi equations}.
\newblock \emph{Transactions of the American Mathematical Society},
  277\penalty0 (1):\penalty0 1--1, 1983.
\newblock \doi{10.1090/S0002-9947-1983-0690039-8}.

\bibitem[Falcone and Ferretti(2013)]{Falcone2013}
Maurizio Falcone and Roberto Ferretti.
\newblock \emph{{Semi-Lagrangian Approximation Schemes for Linear and
  Hamilton-Jacobi Equations}}.
\newblock Society for Industrial and Applied Mathematics, Philadelphia, PA,
  2013.
\newblock ISBN 978-1-61197-304-4.
\newblock \doi{10.1137/1.9781611973051}.

\bibitem[Gonzalez and Rofman(1985)]{Gonzalez1985a}
R.~Gonzalez and E.~Rofman.
\newblock {On Deterministic Control Problems: An Approximation Procedure for
  the Optimal Cost I. The Stationary Problem}.
\newblock \emph{SIAM Journal on Control and Optimization}, 23\penalty0
  (2):\penalty0 242--266, 1985.
\newblock \doi{10.1137/0323018}.

\bibitem[Sethian and Vladimirsky(2003)]{sethian2003ordered}
James~A Sethian and Alexander Vladimirsky.
\newblock Ordered upwind methods for static {H}amilton--{J}acobi equations:
  Theory and algorithms.
\newblock \emph{SIAM Journal on Numerical Analysis}, 41\penalty0 (1):\penalty0
  325--363, 2003.

\bibitem[Bou{\'{e}} and Dupuis(1999)]{Boue1999}
Michelle Bou{\'{e}} and Paul Dupuis.
\newblock {Markov Chain Approximations for Deterministic Control Problems with
  Affine Dynamics and Quadratic Cost in the Control}.
\newblock \emph{SIAM Journal on Numerical Analysis}, 36\penalty0 (3):\penalty0
  667--695, 1999.
\newblock \doi{10.1137/S0036142997323521}.

\bibitem[Zhao(2004)]{Zhao2004}
Hongkai Zhao.
\newblock {A fast sweeping method for Eikonal equations}.
\newblock \emph{Mathematics of Computation}, 74\penalty0 (250):\penalty0
  603--628, 2004.

\bibitem[Qian et~al.(2007)Qian, Zhang, and Zhao]{Qian2007}
Jianliang Qian, Yong‐Tao Zhang, and Hong‐Kai Zhao.
\newblock {Fast Sweeping Methods for Eikonal Equations on Triangular Meshes}.
\newblock \emph{SIAM Journal on Numerical Analysis}, 45\penalty0 (1):\penalty0
  83--107, 2007.

\bibitem[Alton and Mitchell(2012)]{alton2012ordered}
Ken Alton and Ian~M Mitchell.
\newblock An ordered upwind method with precomputed stencil and monotone node
  acceptance for solving static convex {H}amilton-{J}acobi equations.
\newblock \emph{Journal of Scientific Computing}, 51\penalty0 (2):\penalty0
  313--348, 2012.

\bibitem[Mirebeau(2014)]{mirebeau2014efficient}
Jean-Marie Mirebeau.
\newblock Efficient fast marching with {F}insler metrics.
\newblock \emph{Numerische mathematik}, 126\penalty0 (3):\penalty0 515--557,
  2014.

\bibitem[Chacon and Vladimirsky(2012)]{Chacon2012}
Adam Chacon and Alexander Vladimirsky.
\newblock {Fast Two-scale Methods for Eikonal Equations}.
\newblock \emph{SIAM Journal on Scientific Computing}, 34\penalty0
  (2):\penalty0 A547--A578, 2012.

\end{thebibliography}
\end{document}